\documentclass[english,aps,preprint]{revtex4}
\usepackage[T1]{fontenc}
\usepackage[latin9]{inputenc}
\usepackage{array}
\usepackage{multirow}
\usepackage{slashed}
\usepackage{graphicx}

\makeatletter

\providecommand{\tabularnewline}{\\}

\@ifundefined{textcolor}{}
{%
 \definecolor{BLACK}{gray}{0}
 \definecolor{WHITE}{gray}{1}
 \definecolor{RED}{rgb}{1,0,0}
 \definecolor{GREEN}{rgb}{0,1,0}
 \definecolor{BLUE}{rgb}{0,0,1}
 \definecolor{CYAN}{cmyk}{1,0,0,0}
 \definecolor{MAGENTA}{cmyk}{0,1,0,0}
 \definecolor{YELLOW}{cmyk}{0,0,1,0}
}

\usepackage{babel}

\makeatother

\usepackage{babel}
\begin{document}

\title{Doubly Charged Lepton Search Potential of the FCC-Based Energy-Frontier
Electron-Proton Colliders}

\author{A. Ozansoy}
\email{aozansoy@science.ankara.edu.tr}

\selectlanguage{english}%

\affiliation{Ankara University, Faculty of Sciences, Department of Physis, 06100
Tandogan-Ankara, TURKEY }
\begin{abstract}
We search for the doubly charged leptons ($L^{--}$) predicted in
composite models including extended weak isospin multiplets namely,
$I_{W}=1$ and $I_{W}=3/2$ at the Future Circular Collider (FCC)-based
energy-frontier electron-proton colliders with the center-of-mass
energies of $\sqrt{s}=3.46$ TeV, $\sqrt{s}=10$ TeV, and $\sqrt{s}=31.6$
TeV, respectively. We deal with the $e^{-}p\rightarrow L^{--}X\rightarrow e^{-}W^{-}X$
process, calculate the production cross sections, and give the normalized
transverse momentum and pseudorapiditiy distributions of final-state
electron to obtain the kinematical cuts for the discovery. We show
the statistical significance ($SS$) of the expected signal yield as a function of doubly charged lepton mass ($SS-M_{L}$ plots) to attain the doubly charged lepton discovery
mass limits both for the $I_{W}=1$ and $I_{W}=3/2$. It is obtained
that discovery mass limits on the mass of doubly charged lepton for
$I_{W}=1$ ($I_{W}=3/2$) are, $2.21\,(2.73)$ TeV, $5.46\,(8.47)$ TeV, and $12.9\,(20.0)$ TeV
for $\sqrt{s}=3.46$ TeV, $\sqrt{s}=10$ TeV, and $\sqrt{s}=31.6$
TeV, respectively. 
\end{abstract}
\maketitle

\section{introduction }

The spectacular operation of the Large Hadron Collider (LHC) has so
far confirmed the validity of the Standard Model (SM) of particle
physics with great precision. Especially, Higgs boson discovery by
ATLAS and CMS Collaborations at the LHC in 2012 was a great triumph
of the SM \cite{mycitation,key-1}. Nevertheless, there are some issues
that SM gives no explanation such as particle dark matter, neutrino
masses, large number of fundamental particles, lepton-quark symmetry
and fermionic family replication, and it is expected that these issues
will answered at the forthcoming decades by the future high-energy
colliders. Currently, the spectrum of the SM matter particles has
a pattern with three generations listed in growing mass both for lepton
and quark sector. The second and third fermionic families are replicas
of the first family in the context of charge, spin, weak isospin,
color charge but only differ in mass. The fundamental particle inflation
in the SM and family replication are natural indicators for a further
level of substructure. Compositeness is one of the beyond the SM (BSM)
theories that predict a further level of matter constituents called
preons as the ultimate building blocks and known fermions are
composites of them \cite{key-4,key-2,key-3}. A conspicuous consequence
of lepton and quark substructure would be the existence of excited
states \cite{key-5,key-6,key-7,key-8,key-9}. Considering the known
fermions as ground state, spin-1/2 and weak isospin-1/2 excited fermions
are accepted as the lowest radial and orbital excitation by the composite
models. Excited fermions with higher spins take part in composite
models and are considered as higher excitations \cite{key-10,key-11,key-12,key-13,key-14}.

Mostly, excited fermions belonging to weak isospin singlets or doublets,
i.e., $I_{W}=0$ and $I_{W}=1/2$, are studied in detail at various
colliders, so far. Phenomenological studies on spin-1/2 excited leptons
($l^{\star}$) can be found for the lepton and lepton-hadron colliders
in \cite{key-15,key-16,key-17,key-18,key-19,key-20,key-21}, $e\gamma$
and $\gamma\gamma$ colliders in \cite{key-22,key-23,key-24,key-25,key-26},
and hadron colliders in \cite{key-27,key-28,key-29,key-30,key-31,key-32,key-33}.
LHC sets the most stringent bounds on excited leptons and quarks with
spin-1/2. The mass limits were obtained from single production ($pp\rightarrow ll^{\star}X$,
$l=e,\mu,\tau$) at $\sqrt{s}=8$ TeV including contact interactions
in the $l^{\star}$ production and decay mechanism taking into account
that the compositeness scale is equal to excited lepton mass
($\Lambda=m^{\star}$) and $f=f^\prime=1$, where $f$ and $f^\prime$ are the dimensionless couplings determined by the composite dynamics; the ATLAS Collaboration sets
the mass limits as $m_{e^{\star}}>3000$ GeV, $m_{\mu^{\star}}>3000$
GeV, and $m_{\tau^{\star}}>2500$ GeV at the 95\% confidence level
(C.L.) \cite{key-34}. Also, the obtained mass limits for the excited
neutrinos from pair production processes ($pp\rightarrow\nu^{\star}\nu^{\star}X$) were set
as $m_{\nu^{\star}}>1600$ GeV for all types of excited neutrinos
\cite{key-34} and for the excited quarks from single production processes
($pp\rightarrow q^{\star}X$) the mass limit was set as $m_{q^{\star}}>6000$
GeV \cite{key-35}. For the other mass limits and scale limits within
the scope of lepton and quark compositeness searches, see \cite{key-36}.
Very recently, the first search for excited leptons at $\sqrt{s}=13$
TeV is published by the CMS Collaboration \cite{key-37}. Under the
assumption $\Lambda=m^{\star}$, excited electrons and muons are excluded
for masses below 3.9 and 3.8 TeV, respectively, at 95\% C.L. Also,
the best observed limit on the compositeness scale is obtained as
$\Lambda>25$ TeV for both excited electrons and muons for $m^{\star}\sim$ 1.0
TeV. Furthermore, it is shown in \cite{key-38-1} that the effective models for excited fermions violate unitarity in a certain parameter region of the excited fermion mass and compositeness scale. 

In this work, we consider another aspect of compositeness: weak isospin
invariance. From this point of view, usual weak isospin singlets and
doublets are extended to include triplets and quartets ($I_{W}=1$
and $I_{W}=3/2$) \cite{key-38}. Excited states with exotic charges
with $Q=-2e$ for the lepton sector and $Q=5/3e$ and $Q=-4/3e$ for
the quark sector are included in these exotic multiplets. Here we
only concentrate on doubly charged leptons that appearing in $I_{W}=1$
and $I_{W}=3/2$ multiplets. If there is any signal for doubly charged
leptons at future colliders, SM fermionic family structure and replication
could be explained satisfactorily.

In the literature, doubly charged leptonic states appear in type II
seasaw mechanisms \cite{key-39,key-40,key-41}, in models of strong
electroweak symmety breaking \cite{key-42}, in some extensions of
supersymmetric models \cite{key-43,key-44,key-45,key-46,key-47},
in flavor models in warped extra dimensions and in more general models
\cite{key-48,key-49},in string inspired models \cite{key-50}, and in $3-3-1$ models \cite{key-50-1,key-50-2}. Also,
stable doubly charged leptons have been considered as an acceptable
candidate for cold dark matter \cite{key-51}.

Doubly charged lepton phenomenology is investigated so far at the
LHC \cite{key-52,key-53,key-54,key-55,key-56,key-57,key-58,key-59,key-60,key-61,key-62},
at future linear colliders \cite{key-63,key-64,key-65,key-66}, and
at the Large Hadron-electron Collider (LHeC) \cite{key-67}. Doubly
charged leptons related to the second lepton family are investigated at
various possible future muon-proton colliders in \cite{key-68}. Also, the ATLAS
and CMS Collaborations have performed the searches for long-lived
doubly charged states by Drell-Yan-like pair production processes.
The ATLAS Collaboration has excluded long-lived doubly-charged
lepton states masses up to 660 GeV based on the run at $\sqrt{s}=8$
TeV with $L=20.3$ $fb^{-1}$ \cite{key-69} and CMS Collaboration
sets the lower mass limit up to 685 GeV based on the run at $\sqrt{s}=8$
TeV with $L=18.8$ $fb^{-1}$ \cite{key-70}.

LHC is world's largest particle physics laboratory, and it is necessary
to extend its discovery potential and to plan for the colliders after
it. Firstly, a major upgrade of the LHC is High-Luminosity
phase (HL-LHC) \cite{key-71,key-72} with an integrated luminoisty
of 3 $ab^{-1}$ at $\sqrt{s}=14$ TeV and, secondly, a possible further
upgrade of the LHC is High-Energy phase (HE-LHC) \cite{key-73}
with the $27$ TeV center-of mass energy in 2020s.

Future Circular Collider (FCC) project is an exciting and consistent
post-LHC high energy $pp$ collider project at CERN with a center-of-
mass energy of $100$ TeV, and it is supported by European Union within
the Horizon 2020 Framework Programme for Research and Innovation \cite{key-74,key-75}.
Besides the $pp$ option (FCC-hh), FCC includes an electron-positron
collider option (FCC-ee) known as TLEP \cite{key-76,key-77} in the
same tunnel, and also an $ep$ collider option (FCC-eh) providing
the electron beam with an energy of $60$ GeV by an energy recovery
linac (ERL) \cite{key-74}. The FCC-eh would operate concurrently
with the FCC-hh. Same ERL design has been studied in detail as the
main option for the LHeC project
\cite{key-78,key-79}. Concerning ERL that would be positioned inside the
FCC tunnel, energy of the electron beam is limited ($E_{e}<200$ GeV)
due to the large synchroton radiation. To achieve higher electron
beam energies for the $ep$ option of the FCC, linear colliders should
be constructed tangential to the FCC \cite{key-80}. Besides the main
choice of FCC-eh, namely, ERL60, other designs of FCC based $ep$ collider
could be configured using the main parameters of International Linear
Collider (ILC) \cite{key-81} and Plasma Wake Field Accelerator-Linear
Collider (PWFA-LC) \cite{key-82}. A very detailed considerations
on the multi-TeV $ep$ colliders based on FCC and linear colliders
(LC) can be found in \cite{key-80}. Another remarkable and important post-LHC project is Super proton proton Collider (SppC) project which is planned to be built in China with the center-of-mass energy about 70 TeV \cite{key-82-1}. Different options of FCC-based
$ep$ colliders are listed in Table I.

In this work, in Section II we give the basics of extended weak isospin
models and introduce the effective Lagrangians for the gauge interactions
of doubly charged leptons. We consider the production of doubly charged
leptons at future various high-energy $ep$ colliders, show our analysis
to obtain the best cuts for the discovery, and give the obtained mass
limits in Section III, and then, we conclude.

\begin{table}[ht]
\caption{Main parameters of the FCC based $ep$ colliders with the proton beam
energy of $E_{p}=50$ TeV. }

\begin{tabular}{|c|c|c|c|}
\hline 
Collider Name  & $E_{e}$(TeV)  & $\sqrt{s}$ (TeV)  & $L_{int}$($fb^{-1}$per year)\tabularnewline
\hline 
\hline 
ERL60$\otimes$FCC  & 0.06  & 3.46  & 100\tabularnewline
\hline 
ILC$\otimes$FCC  & 0.5  & 10  & 10-100\tabularnewline
\hline 
PWFA-LC$\otimes$FCC  & 5  & 31.6 & 1-10\tabularnewline
\hline 
\end{tabular}
\end{table}

\section{extended weak \i sospin multiplets}

Long before the experimental verification of the existence of quarks
and gluons, strong isospin symmetry allowed to designate the possible
patterns of baryonic and mesonic states and to learn about the properties
of these hadronic states. With the same point of view, using the weak
isospin symmetry arguments, possible fermionic resonances could be
revealed. Thus, without knowing about the dynamics of the fermionic
integral parts (preons) exactly, we could obtain the quantum numbers
of the excited fermionic spectrum. The weak isospin invariance is used to determine the allowed exotic states. SM fermions exist in singlets or
doublets ($I_{W}=0$ or $I_{W}=1/2$) and gauge bosons have $I_{W}=0$
(for photons) or $I_{W}=1$ (for weak bosons), so only $I_{W}\leq3/2$ states
can be allowed. Therefore, usual weak isospin states can be extended to $I_{W}=1$
and $I_{W}=3/2$ states. The details of extended isospin models can
be found in \cite{key-38}. The form of these exotic $I_{W}=1$ and
$I_{W}=3/2$ multiplets are

\begin{equation}
L_{1}=\left(\begin{array}{c}
L^{0}\\
L^{-}\\
L^{--}
\end{array}\right),\qquad L_{3/2}=\left(\begin{array}{c}
L^{+}\\
L^{0}\\
L^{-}\\
L^{--}
\end{array}\right)
\end{equation}

and similar for the antiparticles. These multiplets can be arranged
for all flavor of leptons. Also, exotic multiplets with $I_{W}=1$
and $I_{W}=3/2$ exist in the quark sector. 

To attain the decay widths and production cross sections, we have to
specify the doubly charged lepton couplings to SM leptons and
gauge bosons. Due to the lack of knowledge about the explicit dynamics
of preons, we use the effective Lagrangian method. Since all the gauge
fields have $Y=0$ weak hypercharge, a certain exotic multiplet couples
through the gauge fields to a SM multiplet with the same $Y$. According
to the well-known Gell-Mann - Nishijima formula ($Q=I_{3}+\frac{Y}{2}$), exotic multiplets $I_{W}=1$ has $Y=-2$ and $I_{W}=3/2$ has $Y=-1$, so $L^{--}$ from $I_{W}=1$ couples to SM right-handed leptons (singlets) and
$L^{--}$ from $I_{W}=3/2$ couples to SM left-handed leptons (doublets).
To assure the current conservation, the couplings have to be of anomalous
magnetic moment type. The only contribution that involves both $I_{W}=1$
and $I_{W}=3/2$ comes from the isovector current. Thus, doubly charged
leptons can couple to SM leptons only via $W^{\pm}$ gauge bosons.
Relevant gauge-mediated interaction Lagrangians which are made of
dimension five operators to describe the interactions between a doubly
charged lepton, a SM lepton and $W$ boson for the exotic multiples
are given by

\begin{equation}
\mathcal{L}_{GM}^{(1)}=i\frac{gf_{1}}{\Lambda}(\bar{L}\sigma_{\mu\nu}\partial^{\nu}W^{\mu}\frac{1+\gamma_{5}}{2}\ell)+h.c
\end{equation}

\begin{equation}
\mathcal{L}_{GM}^{(3/2)}=i\frac{gf_{3/2}}{\Lambda}(\bar{L}\sigma_{\mu\nu}\partial^{\nu}W^{\mu}\frac{1-\gamma_{5}}{2}\ell)+h.c
\end{equation}

Here, $g$ is the $SU(2)$ coupling and equal to $g_{e}/sin\theta_{W}$
where $g_{e}=\sqrt{4\pi\alpha}$ , $\theta_{W}$ is weak mixing
angle and $\alpha$ is the fine structure constant, $\sigma_{\mu\nu}$
is the antisymmetric tensor being $\sigma_{\mu\nu}=\frac{i}{2}(\gamma_{\mu}\gamma_{\nu}-\gamma_{\nu}\gamma_{\mu})$,
$\Lambda$ is the compositeness scale, $f_{1}$ and $f_{3/2}$
are the couplings which are responsible for the effective interactions
of $I_{W}=1$ and $I_{W}=3/2$ multiplets, respectively. $L$ denotes
the doubly charhed lepton, $l$ denotes the SM lepton. The vertex factors can be inferrred
from Eq.2 and Eq.3 as

\begin{equation}
\Theta_{\mu}^{(i)}=\frac{g_{e}f_{i}}{4\Lambda sin\theta_{W}}\left(\gamma_{\mu}\slashed{q}-\slashed{q}\gamma_{\mu}\right)\left(1\mp\gamma_{5}\right)\qquad i=1,3/2
\end{equation}
where $\slashed{q}=q^{\nu}\gamma_{\nu}$ and $q^{\nu}$ is the four-momentum
of the gauge field. In Eq.4, $+$ is for $i=1$ and $-$ is for $i=3/2$. Due to the fact that the only contribution to the interaction Lagrangian comes from isovector current, $L^{--}$ has only one decay mode $L^{--}\rightarrow W^{-}l^{-}$.
Neglecting SM lepton mass, the analytical expression for the decay
width of doubly charged lepton is

\begin{equation}
\Gamma(L^{--}\rightarrow W^{-}l^{-})=\left(\frac{f}{sin\theta_{W}}\right)^{2}\alpha\left(\frac{M_{L}^{3}}{8\Lambda^{2}}\right)\left(1-\frac{m_{W}^{2}}{M_{L}^{2}}\right)^{2}\left(2+\frac{m_{W}^{2}}{M_{L}^{2}}\right)
\end{equation}

and Eq.5 has the same form both for $I_{W}=1$ and $I_{W}=3/2$ as
we set $f_{1}=f_{3/2}=f$. In Figure 1, we plot the decay width of doubly charged lepton as a function of its mass for three different values of $\Lambda$. Under the considerations $\Lambda=M_L$
and $m_{W}\ll M_{L}$, Eq.5 suggests that doubly charged lepton decay
width increases linearly with mass for a specific value of $f$ . 

\begin{figure}[ht]
\includegraphics[scale=0.8]{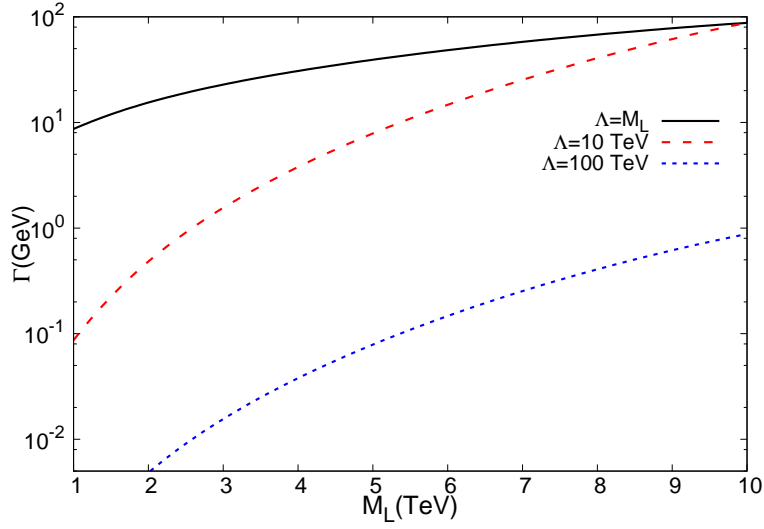}

\caption{Decay width of doubly charged leptons for $\Lambda=M_{L}$, $\Lambda=10$
TeV and $\Lambda=100$ TeV.}
\end{figure}

\section{Doubly charged lepton production at future $ep$ colliders}

Doubly charged leptons can be produced singly via the process $e^{-}p\rightarrow L^{--}X$.
Feynman diagrams for the subprocesses $e^{-}q(\bar{\bar{q^{\prime}})\rightarrow L^{--}q^{\prime}(\bar{q)}}$
are shown in Figure 2.

\begin{figure}[!ht]
\includegraphics[scale=0.8]{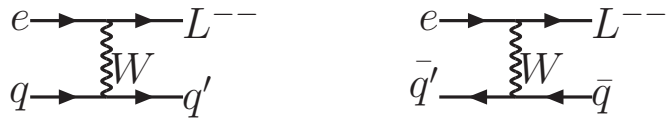}

\caption{Feynman diagrams responsible for the subprocess $e^{-}q\rightarrow L^{--}q^{\prime}$
(left panel) and $e^{-}\bar{q^{\prime}}\rightarrow L^{--}\bar{q}$ (right
panel). }
\end{figure}

Neglecting SM lepton and quark masses, we find the analytical expression
of differantial cross section for taking into account $I_{W}=1$ for
the subprocess $e^{-}q\rightarrow L^{--}q^{\prime}$ is
\begin{equation}
\frac{d\hat{\sigma}}{dt}_{(eq\rightarrow L^{--}q^{\prime)}}=\frac{f_{1}^{2}g^{4}((s-M_{L}^{2})(M_{L}^{2}-s-t)\:t\:|V_{qq^{\prime}}|^{2})}{32\Lambda^{2}\pi s^{2}(m_{W}^{2}-t)^{2}}
\end{equation}

and for the subprocess $e^{-}\bar{q^{\prime}}\rightarrow L^{--}\bar{q}$
is 

\begin{equation}
\frac{d\hat{\sigma}}{dt}_{(e\bar{q^{\prime}}\rightarrow L^{--}\bar{q})}=\frac{-f_{1}^{2}g^{4}((s+t)\:t\:|V_{qq^{\prime}}|^{2})}{32\Lambda^{2}\pi s(m_{W}^{2}-t)^{2}}
\end{equation}

Changing $f_{1}\rightarrow f_{3/2}$ Eq.6 is valid for $e^{-}\bar{q^{\prime}}\rightarrow L^{--}\bar{q}$
and Eq.7 is valid for $e^{-}q\rightarrow L^{--}q^{\prime}$ for $I_{W}=3/2$.
We inserted doubly charged lepton interaction vertices given in Eq.4
into the well-known high-energy physics simmulation programme CalcHEP
\cite{key-83,key-84,key-85} and used it for our calculations. 

Total production cross section for the process $e^{-}p\rightarrow L^{--}X$
both for $I_{W}=1$ and $I_{W}=3/2$ as a function of doubly charged
lepton mass is shown in Figure 3 for taking into account $\Lambda=M_{L}$ (left panel) and $\Lambda=100$ TeV (right panel). We use CTEQ6L parton
distribution function \cite{key-86}. As seen from Figure 3, total
cross sections for the doubly charged leptons for $I_{W}=3/2$ are
slightly larger than the ones for $I_{W}=1$. This result is due to
the contribution of valence quarks in the initial state when $L^{--}$
is being produced.

\begin{figure}[!ht]
\includegraphics[scale=0.7]{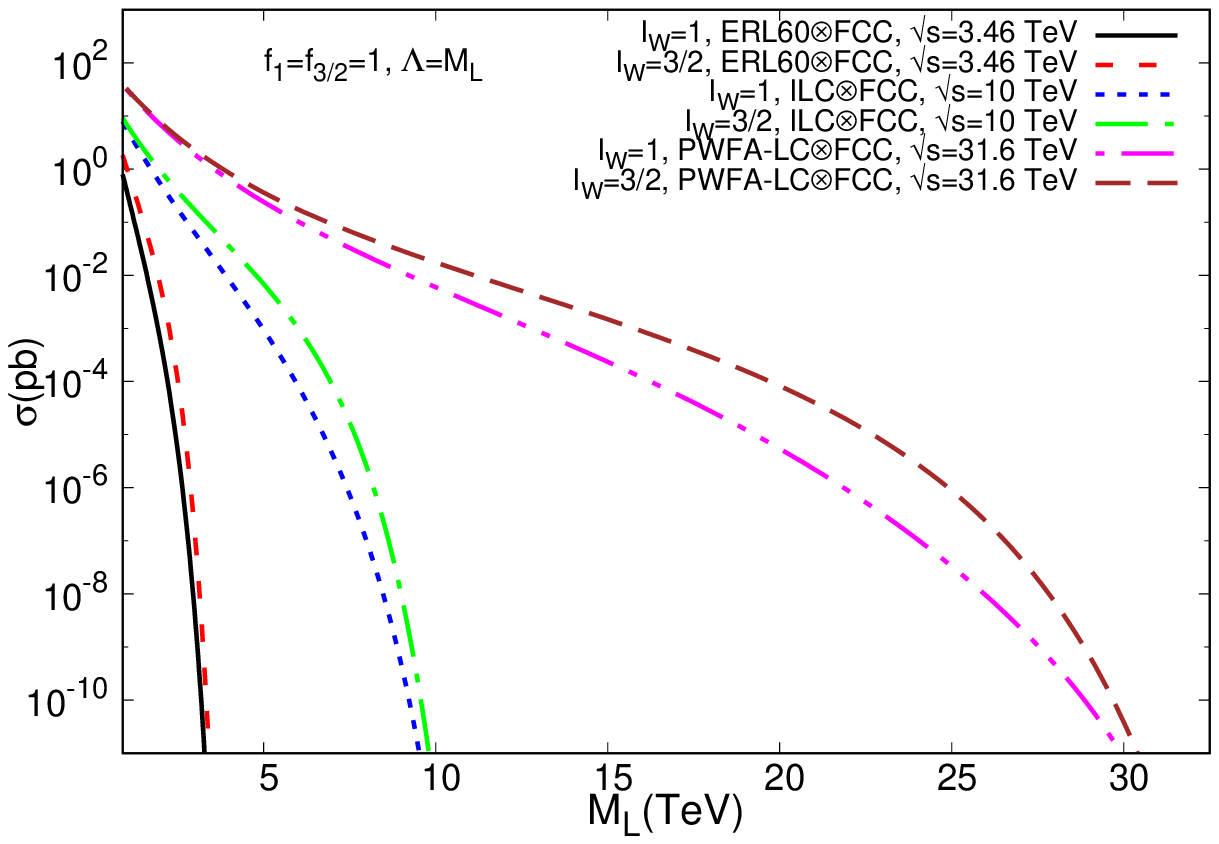}\includegraphics[scale=0.7]{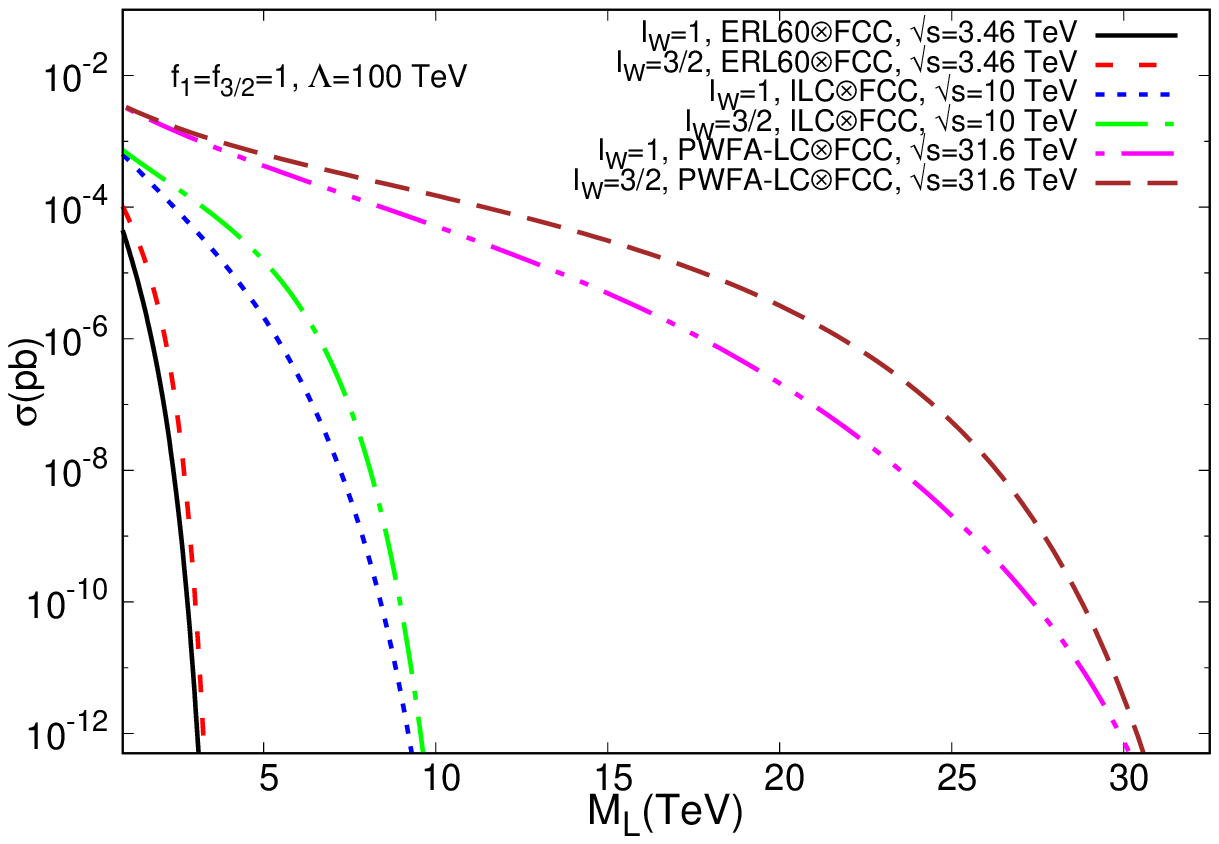}

\caption{Production cross sections for the single production of doubly charged
leptons at future $ep$ colliders for $\Lambda=M_{L}$(left panel)
and $\Lambda=100$ TeV (right panel). }
\end{figure}

Taking into account the decay of $L^{--}$, we consider the kinematical
distributions for the process $e^{-}q({\bar{q^{\prime}}})\rightarrow e^{-}W^{-}q^{\prime}({\bar{q}})$.
Respecting lepton number conservation we only deal with the doubly
charged leptons related to the first generation.

Since design studies are ongoing for an appropriate detector for the $ep$ colliders considered in this work, our analysis is at the parton level. 

We impose the basic cuts for the final-state electron and quarks as

\begin{equation}
p_{T}^{e}>20\:GeV,\:p_{T}^{j}>30\:GeV
\end{equation}

After appyling basic cuts, SM cross sections are $\sigma_{B}=4.04$
pb, $\sigma_{B}=17.52$ pb, and $\sigma_{B}=67.99$ pb for $\sqrt{s}=3.46,\:10,$
and $31.6$ TeV, respectively. To reveal a clear signal, it is very
important to determine the most appropriate cuts. After appliying the basic cuts, we plot the normalized
transverse momentum (in Figures 4 and 5) and normalized pseudorapidity
(in Figures 6 and 7) distributions of final state electron originated
by the $L^{--}$. These distributions
exhibit the same characteristic for $I_{W}=1$ and $I_{W}=3/2$. 

From the normalized $p_{T}$ distributions it is inferred that doubly
charged leptons have high transverse momentum which shows a peak around
$M_{L}/2$ in their distributions. From the normalized $\eta$ distributions
of electron, it is seen that the electrons are in a backward direction, consequently
$L^{--}$ is produced in the backward direction. As the center-of-mass
energy of the collider increases, normalized $\eta$ distributions
become more symmetric. 

Examining normalized $p_{T}$ and $\eta$ distributions we extract
the discovery cuts for the final state electron. We choose the suitable
regions where we eliminate most of the background while not losing most of the siganl. Our results are summarized in Table
II.

\begin{figure}[!ht]
\includegraphics[scale=0.7]{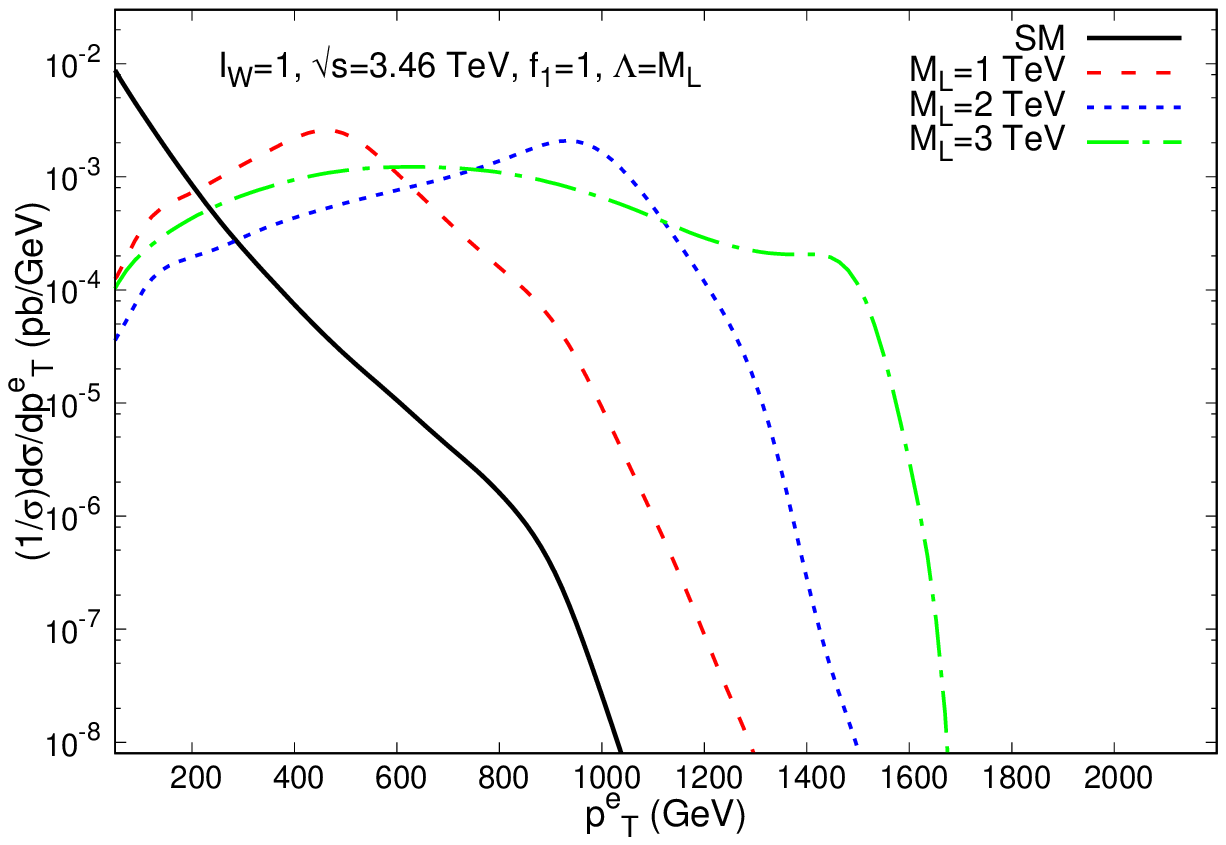}\includegraphics[scale=0.7]{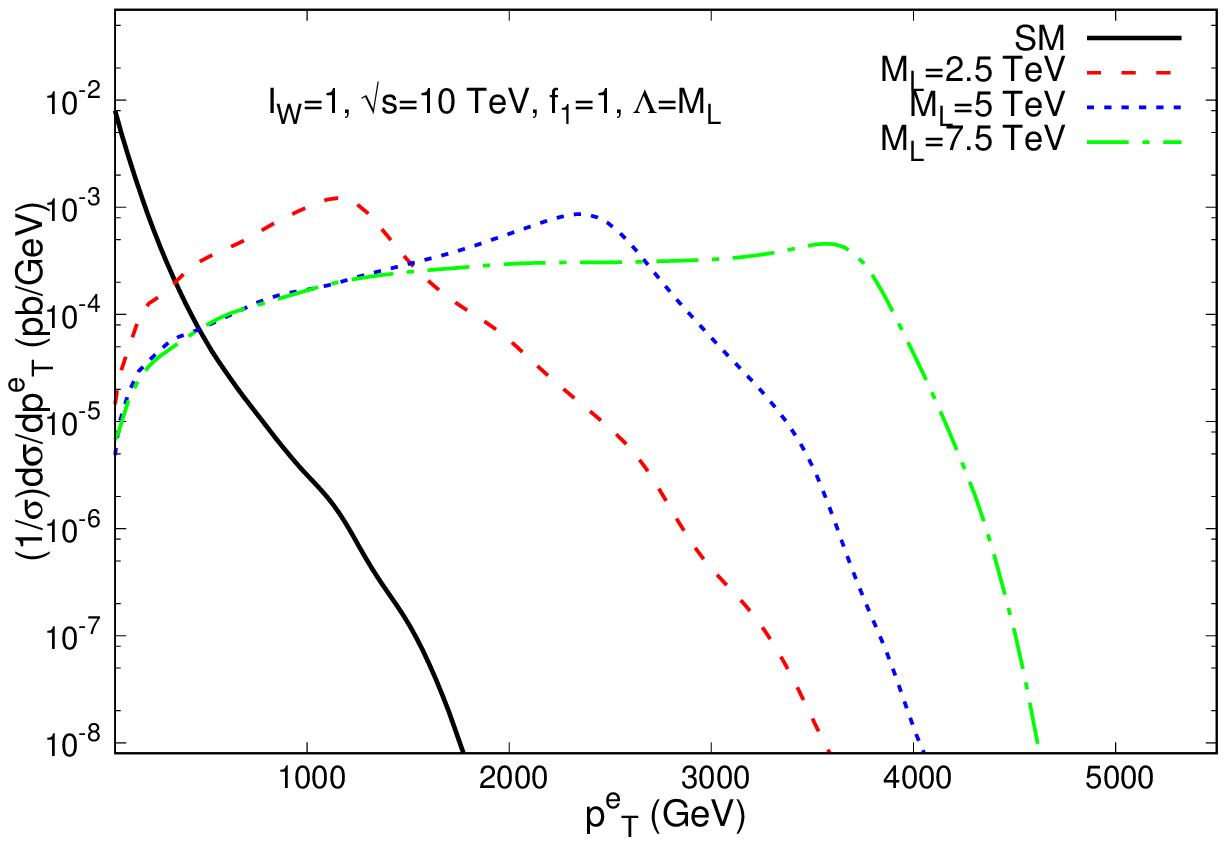}

\includegraphics[scale=0.7]{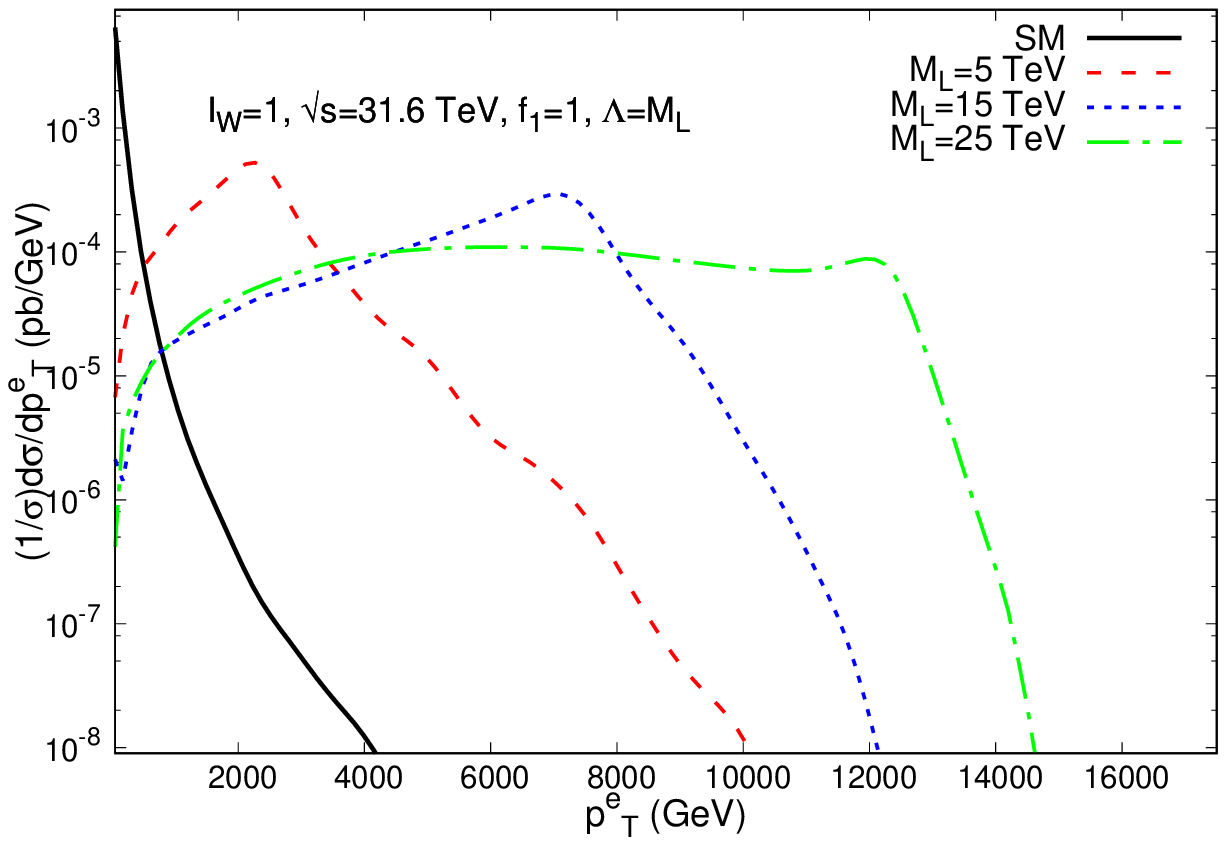}

\caption{Normalized $p_{T}$ distributions of the final state electron for the
$I_{W}=1$ multiplet for $f_{1}=1$ and $\Lambda=M_{L}$ for various
$ep$ colliders. }
\end{figure}

\begin{figure}[!ht]
\includegraphics[scale=0.7]{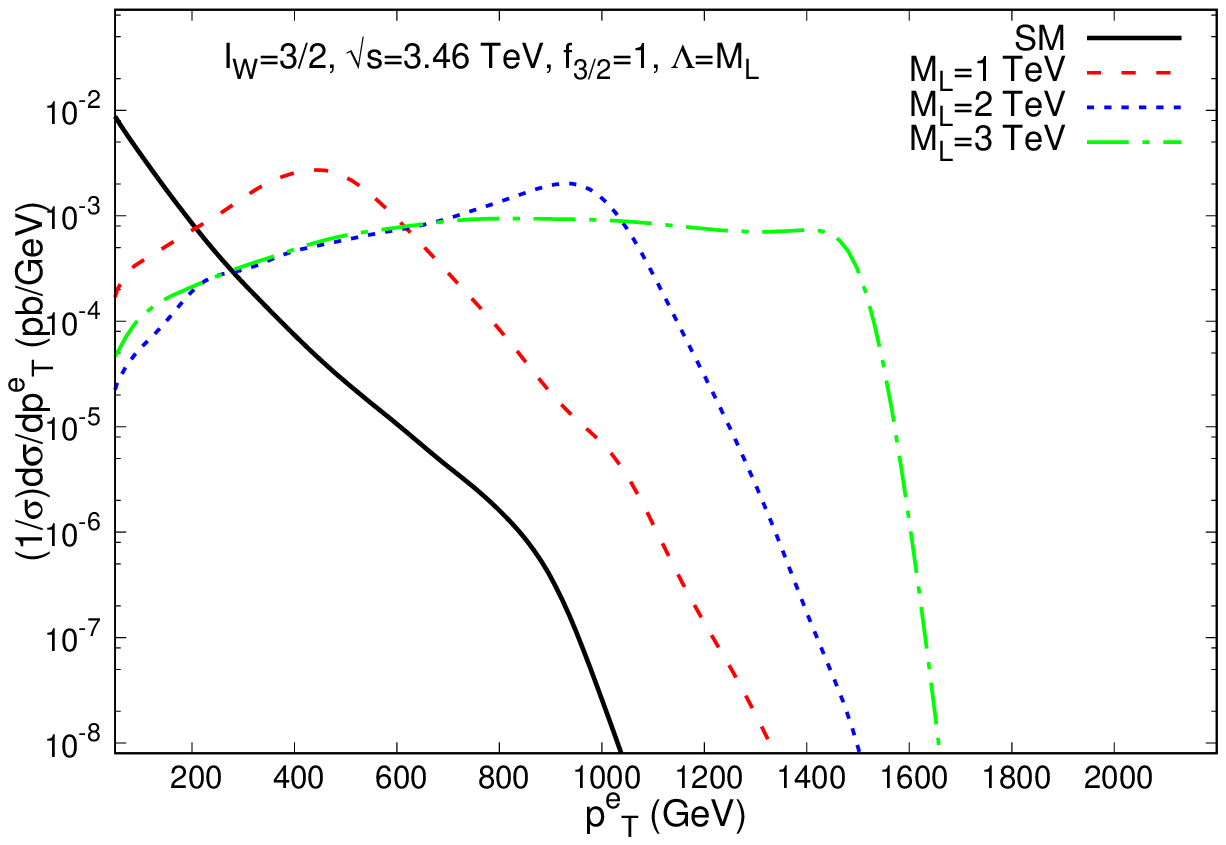}\includegraphics[scale=0.7]{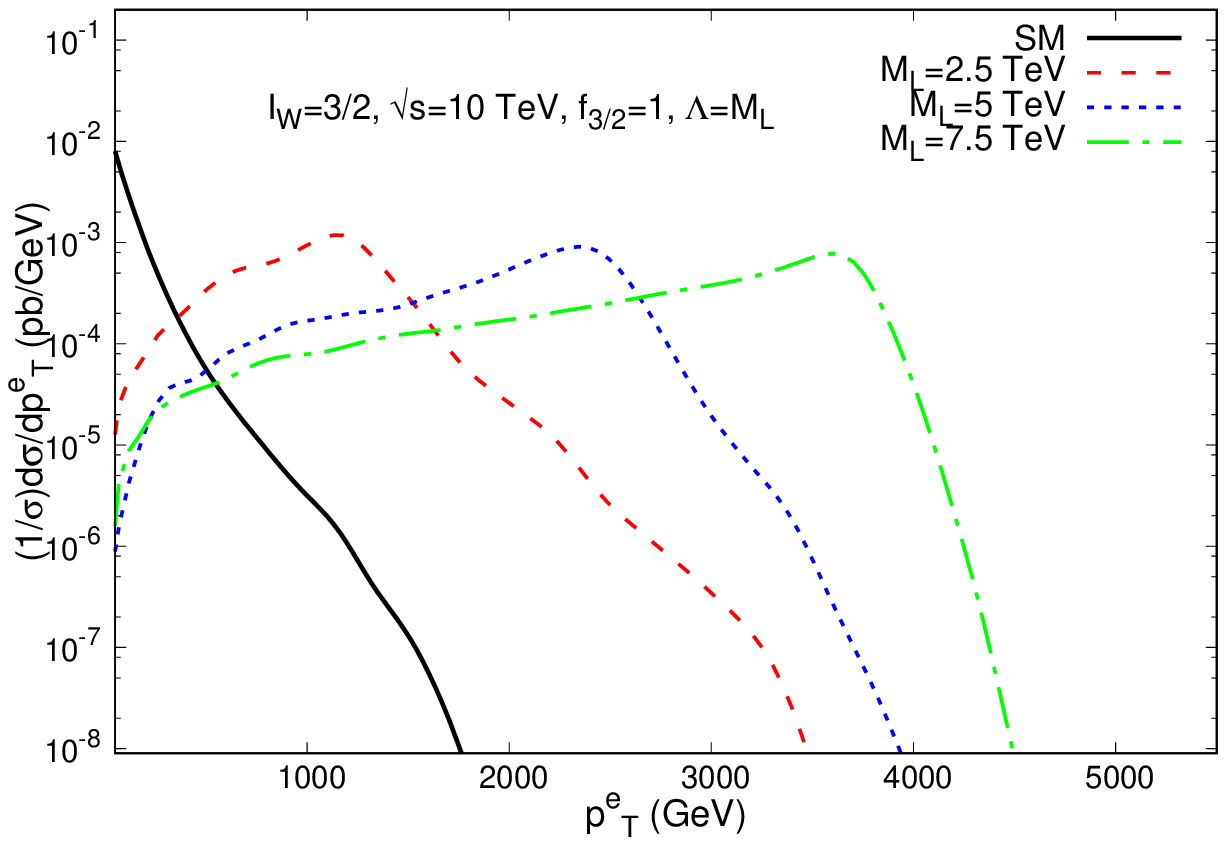}

\includegraphics[scale=0.7]{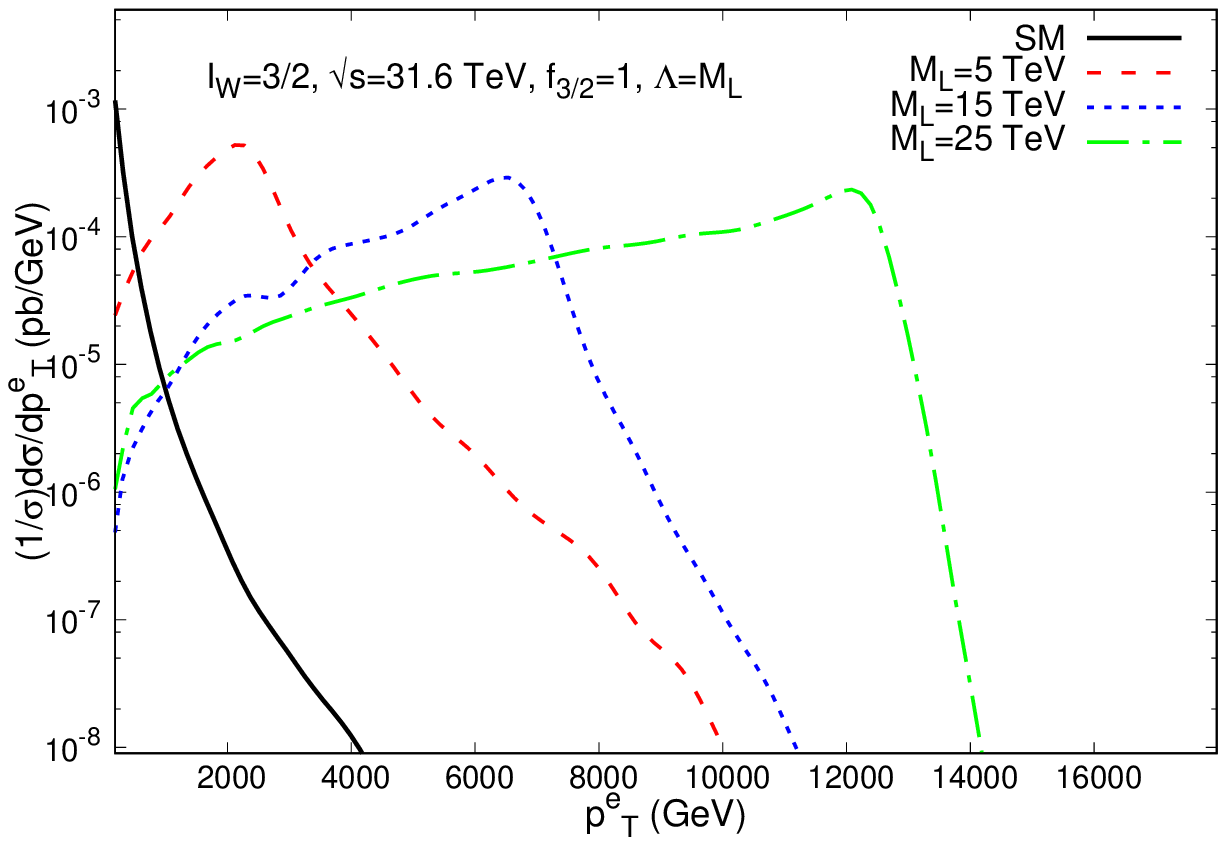}

\caption{Normalized $p_{T}$ distributions of the final state electron for the
$I_{W}=3/2$ multiplet for $f_{3/2}=1$ and $\Lambda=M_{L}$ for various
$ep$ colliders. }
\end{figure}

\begin{figure}[!ht]
\includegraphics[scale=0.7]{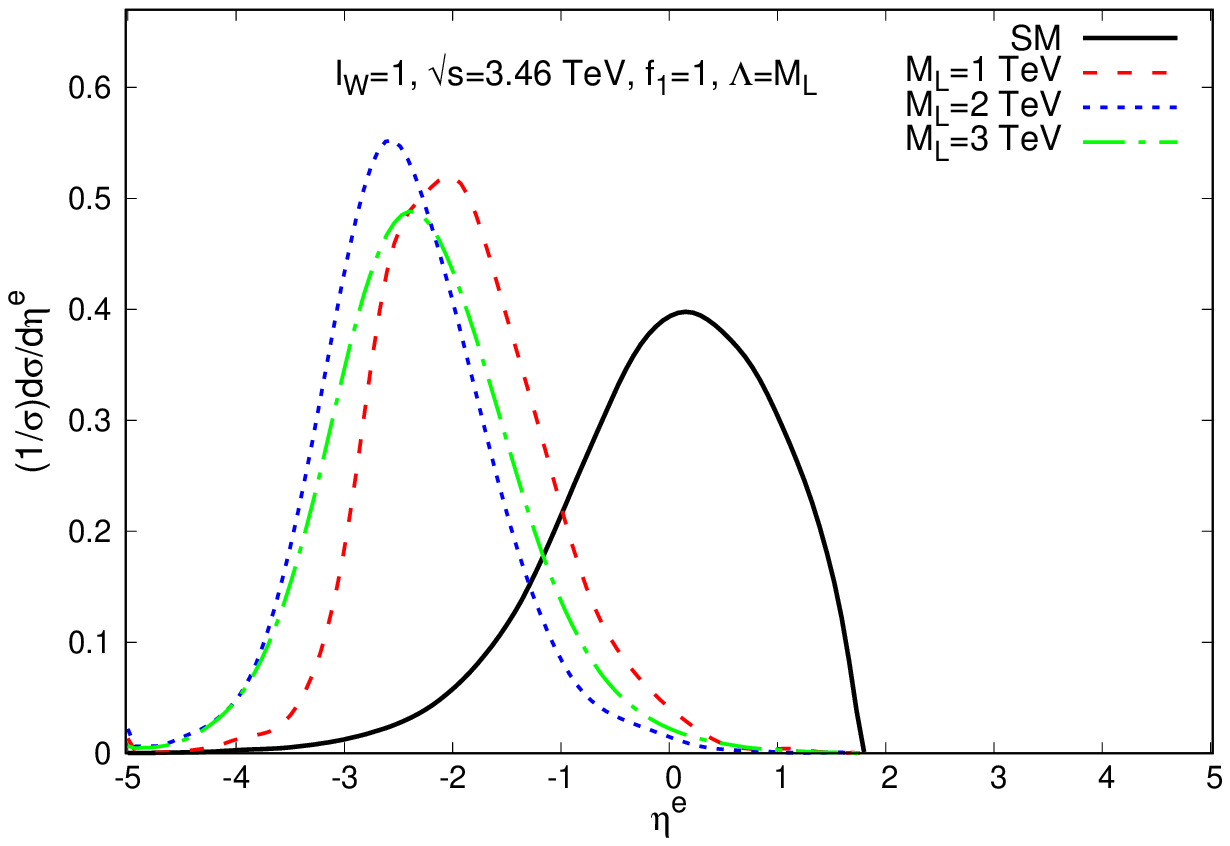}\includegraphics[scale=0.7]{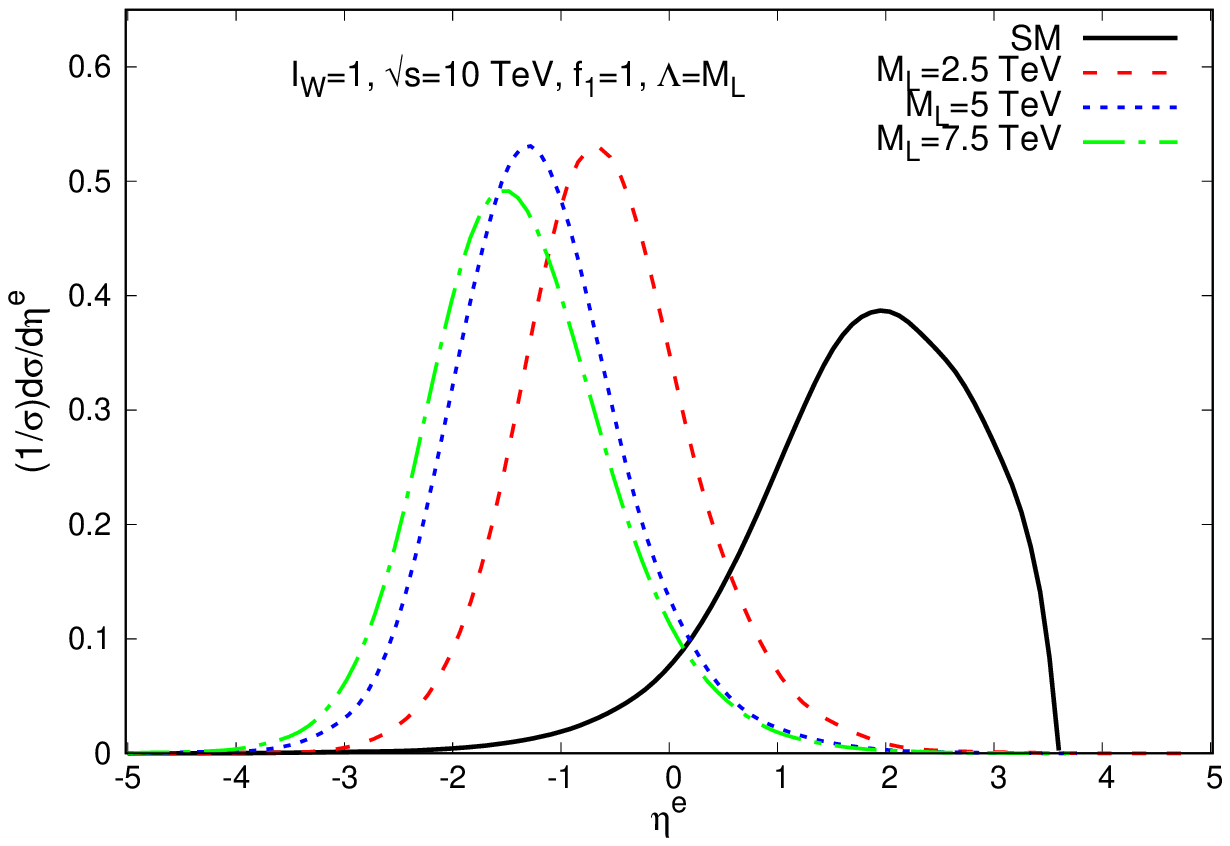}

\includegraphics[scale=0.7]{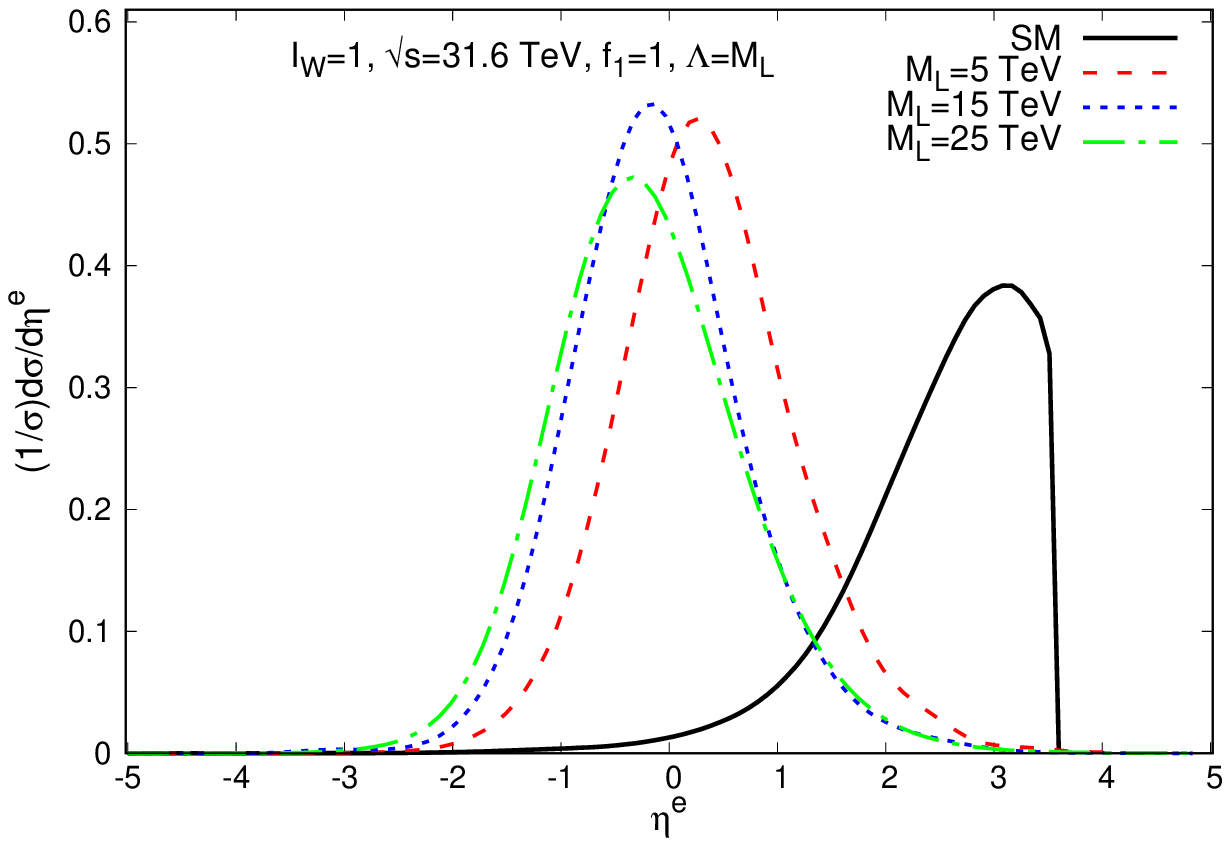}

\caption{Normalized $\eta$ distribution of the final state electron for the
$I_{W}=1$ multiplet for $f_{1}=1$ and $\Lambda=M_{L}$ for various
$ep$ colliders. }
\end{figure}

\begin{figure}[!ht]
\includegraphics[scale=0.7]{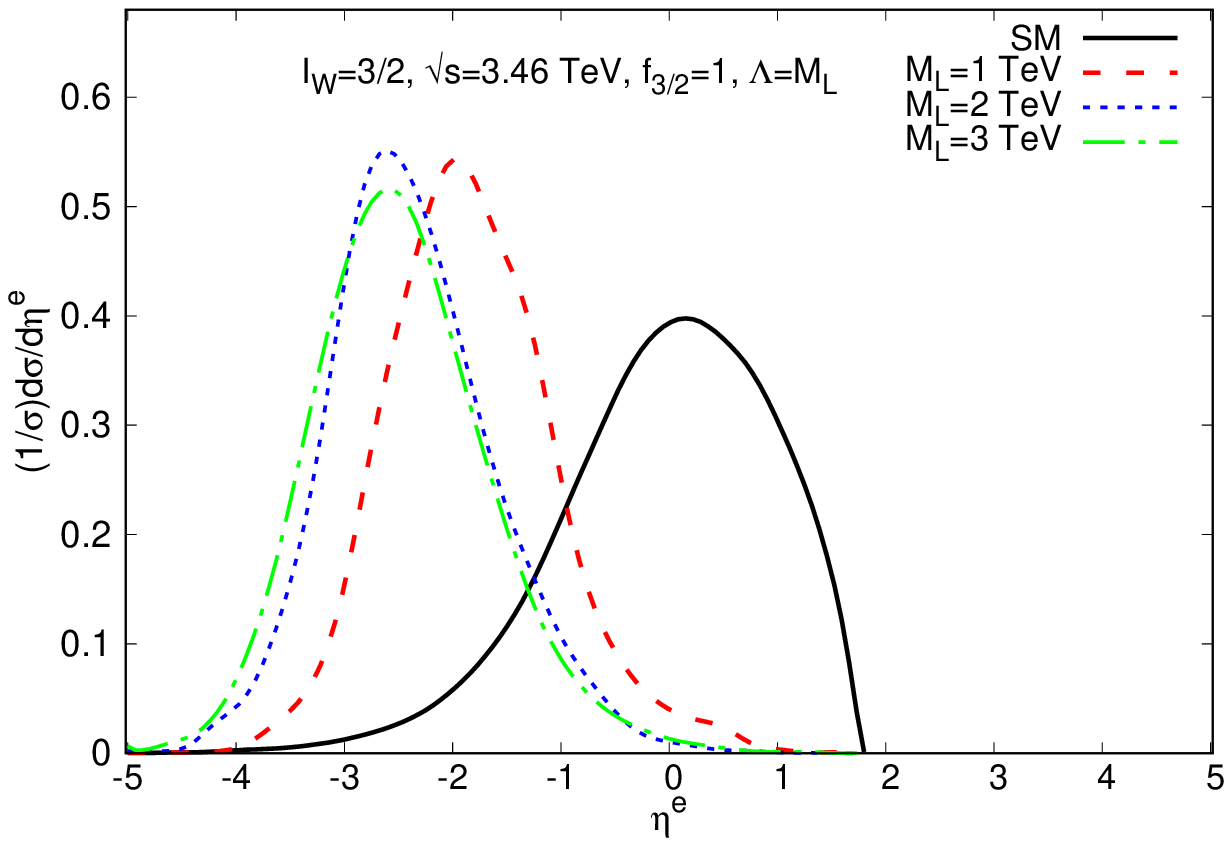}\includegraphics[scale=0.7]{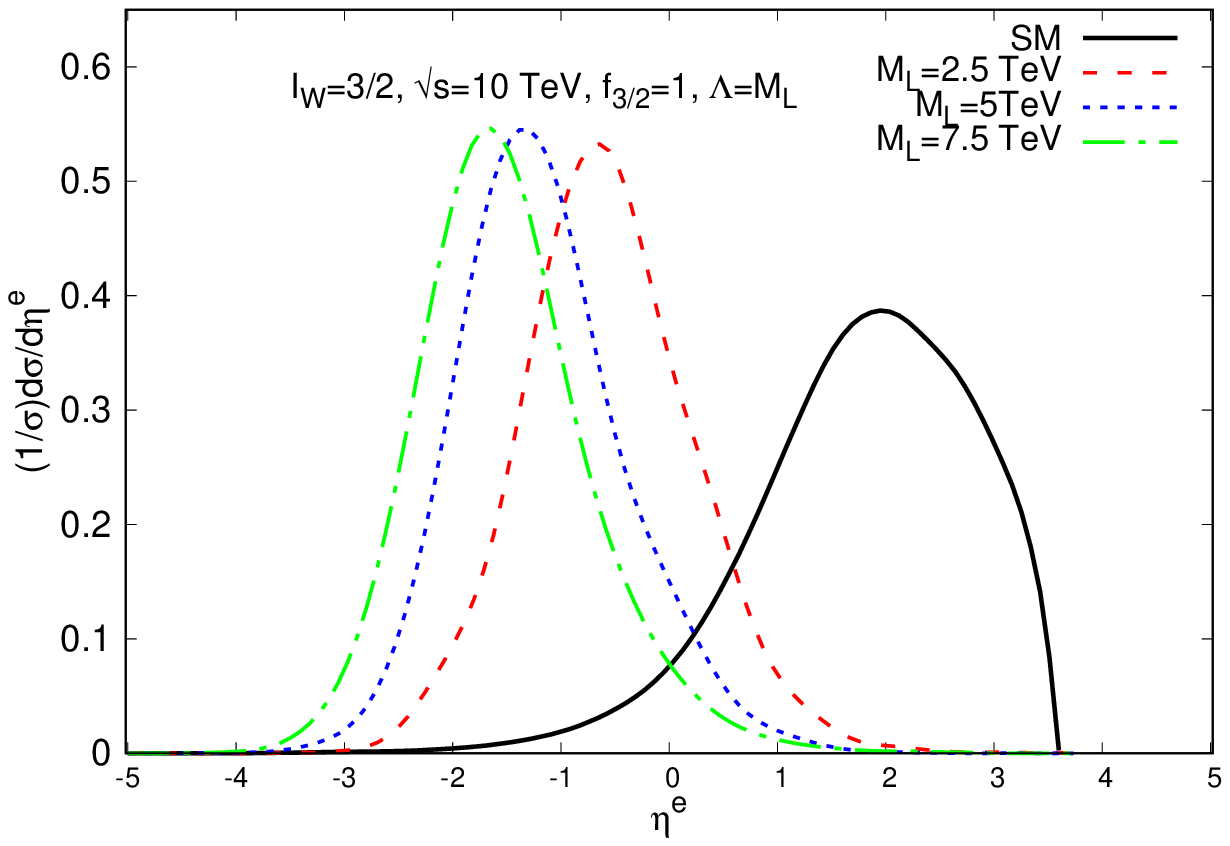}

\includegraphics[scale=0.7]{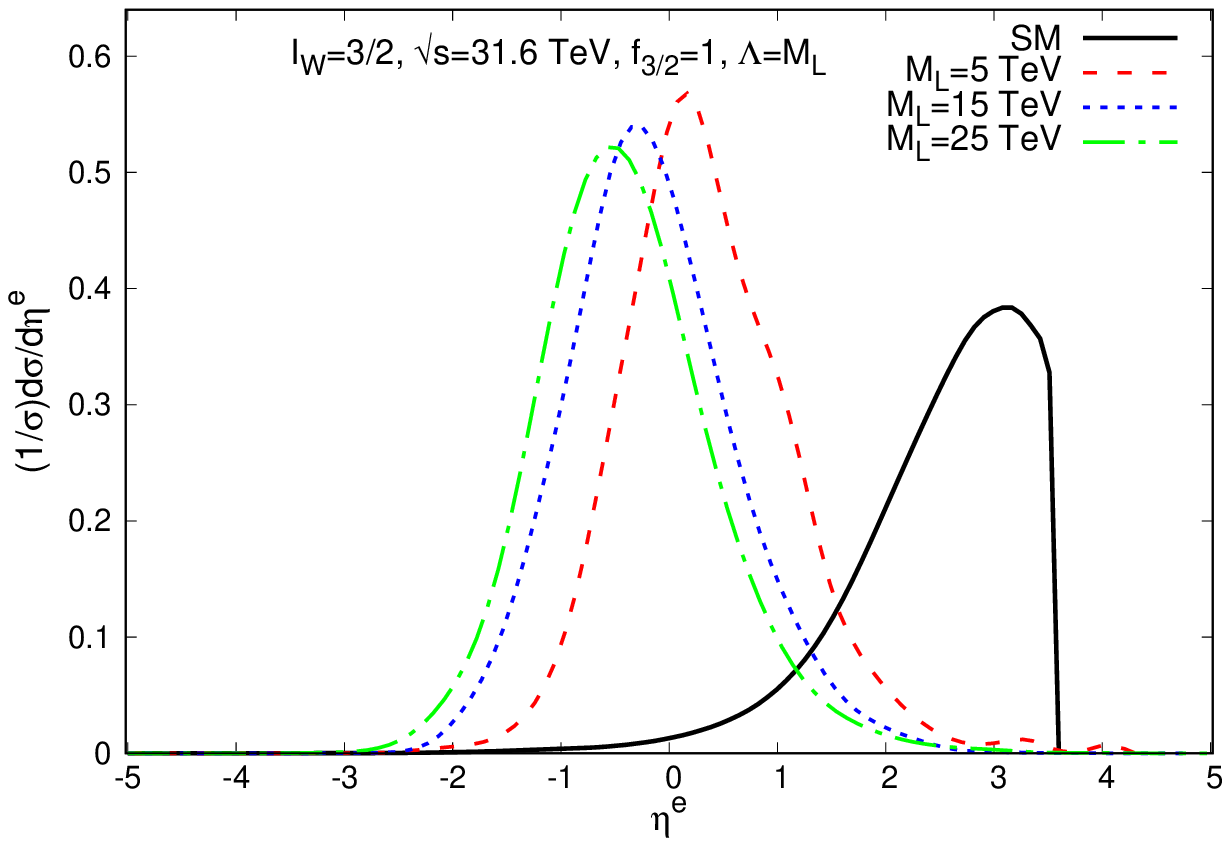}

\caption{Normalized $\eta$ distribution of the final state electron for the
$I_{W}=3/2$ multiplet for $f_{3/2}=1$ and $\Lambda=M_{L}$ for various
$ep$ colliders. }
\end{figure}

\begin{table}[!ht]
\caption{Discovery cuts. }

\begin{tabular}{|c|c|c|c|}
\hline 
\multicolumn{1}{|c|}{} & ERL60$\otimes$FCC  & ILC$\otimes$FCC  & PWFA-LC$\otimes$FCC\tabularnewline
\hline 
\hline 
\multirow{2}{*}{$I_{W}=1$} & $p_{T}^{e}>200$ GeV  & $p_{T}^{e}>340$ GeV  & $p_{T}^{e}>500$ GeV\tabularnewline
\cline{2-4} 
 & $-4<\eta^{e}<-1.1$  & $-3.3<\eta^{e}<0.5$  & $-2.1<\eta^{e}<1.5$\tabularnewline
\hline 
\multirow{2}{*}{$I_{W}=3/2$} & $p_{T}^{e}>210$ GeV  & $p_{T}^{e}>350$ GeV  & $p_{T}^{e}>530$ GeV\tabularnewline
\cline{2-4} 
 & $-4<\eta^{e}<-1$.1  & $-3.3<\eta^{e}<0.5$  & $-2.1<\eta^{e}<1.5$\tabularnewline
\hline 
\end{tabular}
\end{table}

To distinguish the signal and the background, we also imply an invariant
mass cut on $e^{-}W^{-}$ system for the mass intervals (we have selected
the events within the mass intervals).

\begin{equation}
M_{L}-2\Gamma_{L}<M_{eW}<M_{L}+2\Gamma_{L},
\end{equation}

where $\Gamma_{L}$ is the decay width of the doubly charged lepton
for a given value of $M_{L}$. By carrying out the invariant mass
cut, the background cross sections are rather suppressed. 

The final-state signatures obtained from the decays of doubly charged
lepton and $W$ boson are given in Table III. We choose hadronic decay
mode of $W$ boson, $W\rightarrow jj.$ 

\begin{table}[!ht]
\caption{Final states for the doubly charged lepton production at $ep$ colliders.}

\begin{tabular}{|c|c|c|}
\hline 
$L^{--}$decay mode & $W$- boson decay mode  & Final state\tabularnewline
\hline 
\hline 
\multirow{2}{*}{$L^{--}\rightarrow l^{-}W^{-}$} & Leptonic ($W^{-}\rightarrow l^{-}\nu_{l}$)  & $l^{-}(l^{-}\nu_{l})j$ (Same-sign leptons+jet+MET\tabularnewline
\cline{2-3} 
 & Hadronic ($W^{-}\rightarrow2j$)  & $l^{-}(j\,j)\,j$ (Single lepton+3 jet)\tabularnewline
\hline 
\end{tabular}
\end{table}

After implying discovery cuts presented for the final state electron in Table II, we plot the invariant
mass distribution of $e^{-}jj$ system in Figures 8 and 9 for $I_{W}=1$
and $I_{W}=3/2$, respectively. 

\begin{figure}[!htb]
\includegraphics[scale=0.7]{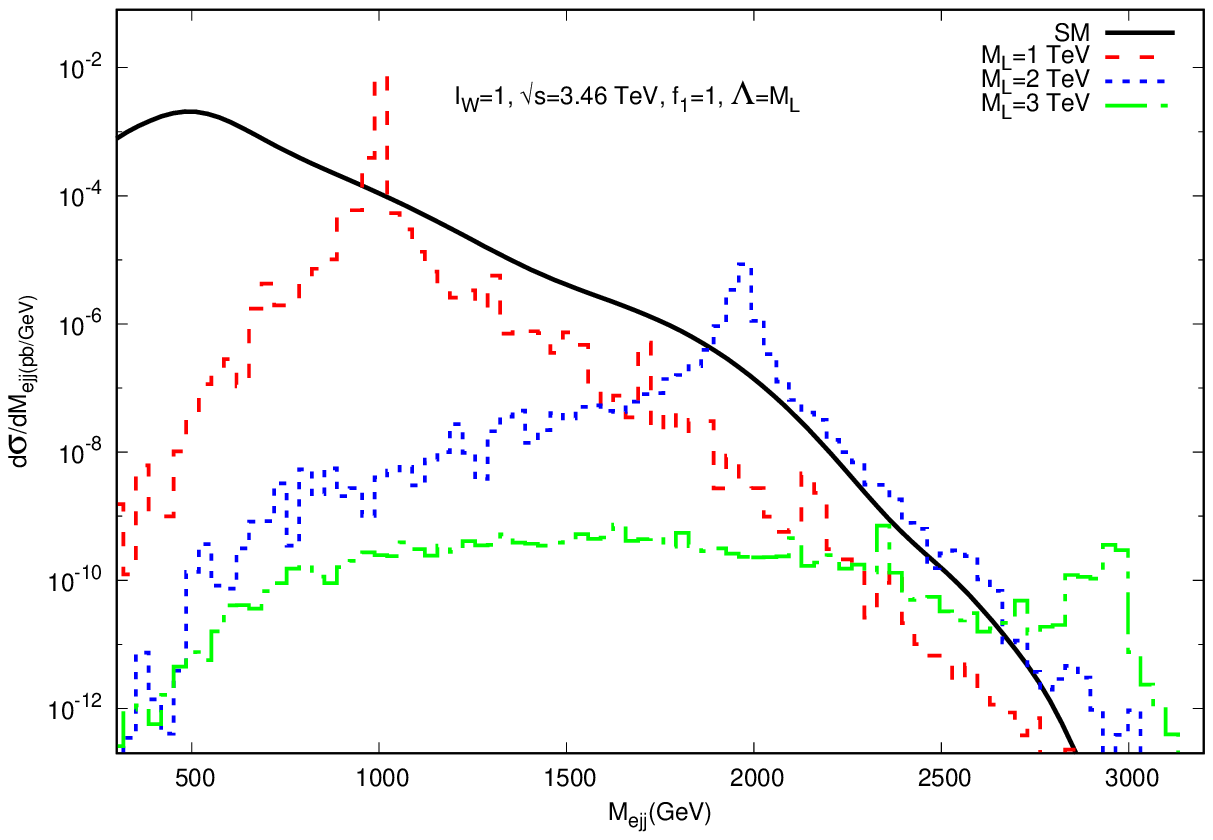}\includegraphics[scale=0.7]{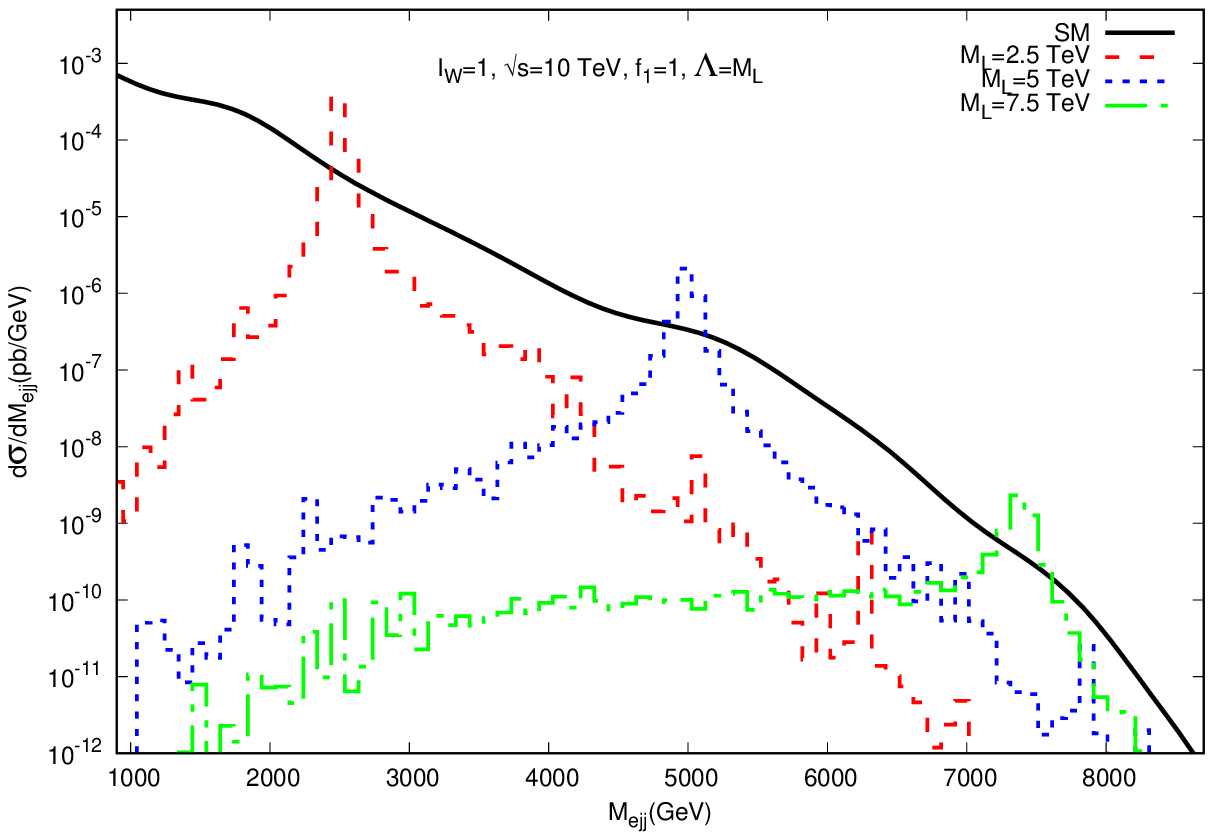}

\includegraphics[scale=0.7]{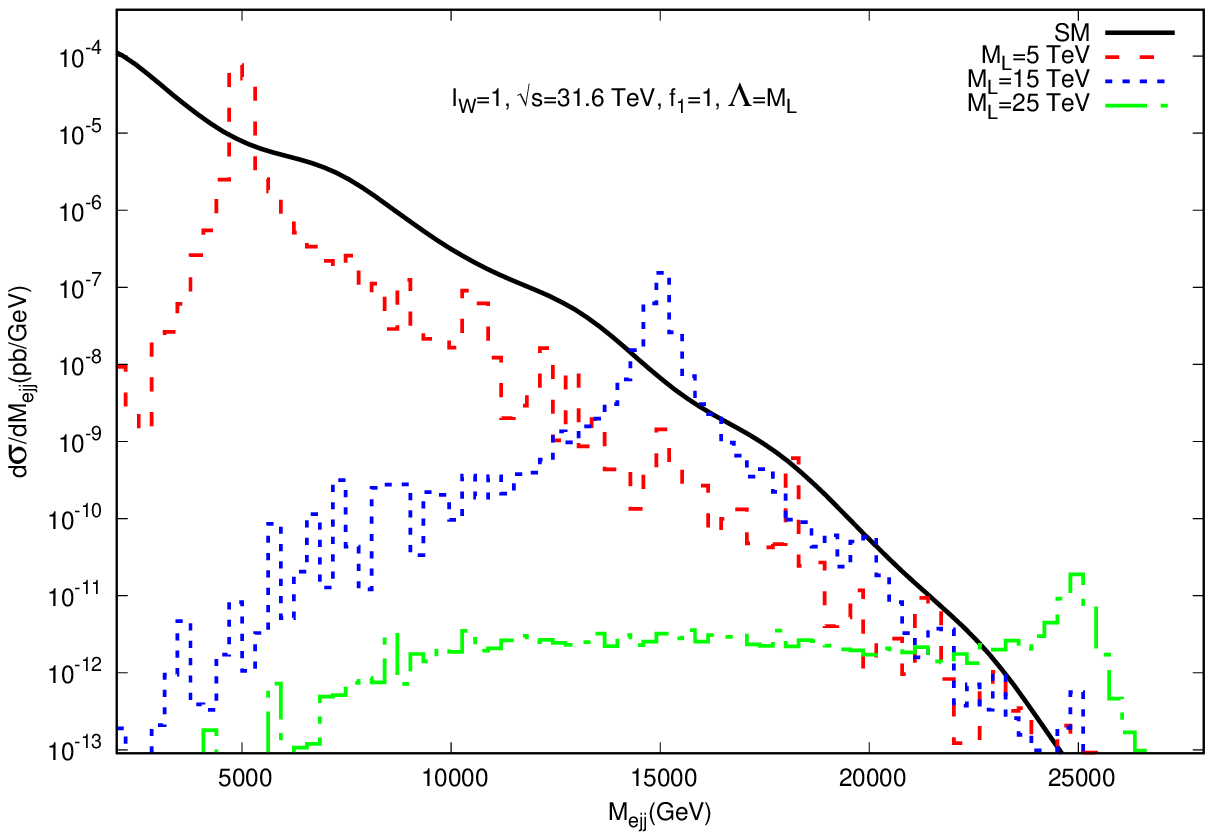}

\caption{Invariant mass distribution of $ejj$ system for $I_{W}=1$ after
the discovery cuts.}
\end{figure}

\begin{figure}[!htb]
\includegraphics[scale=0.7]{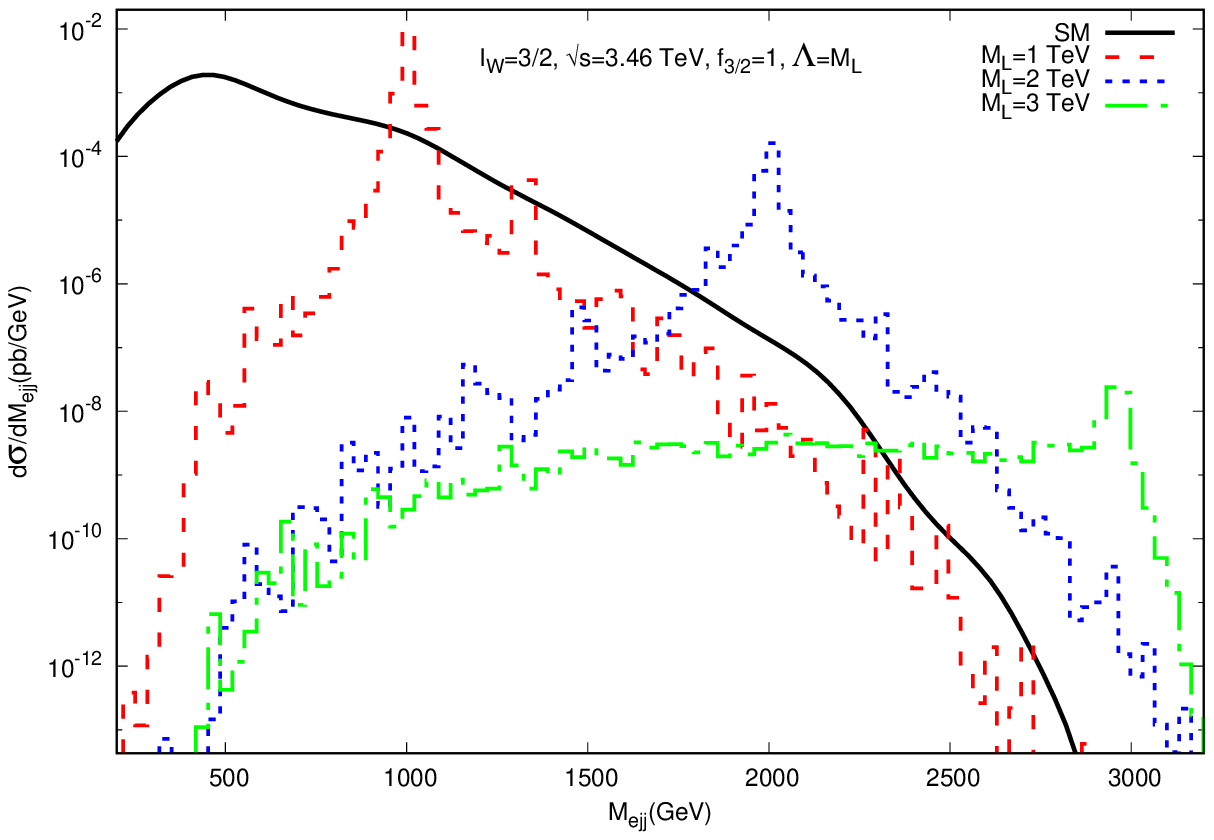}\includegraphics[scale=0.7]{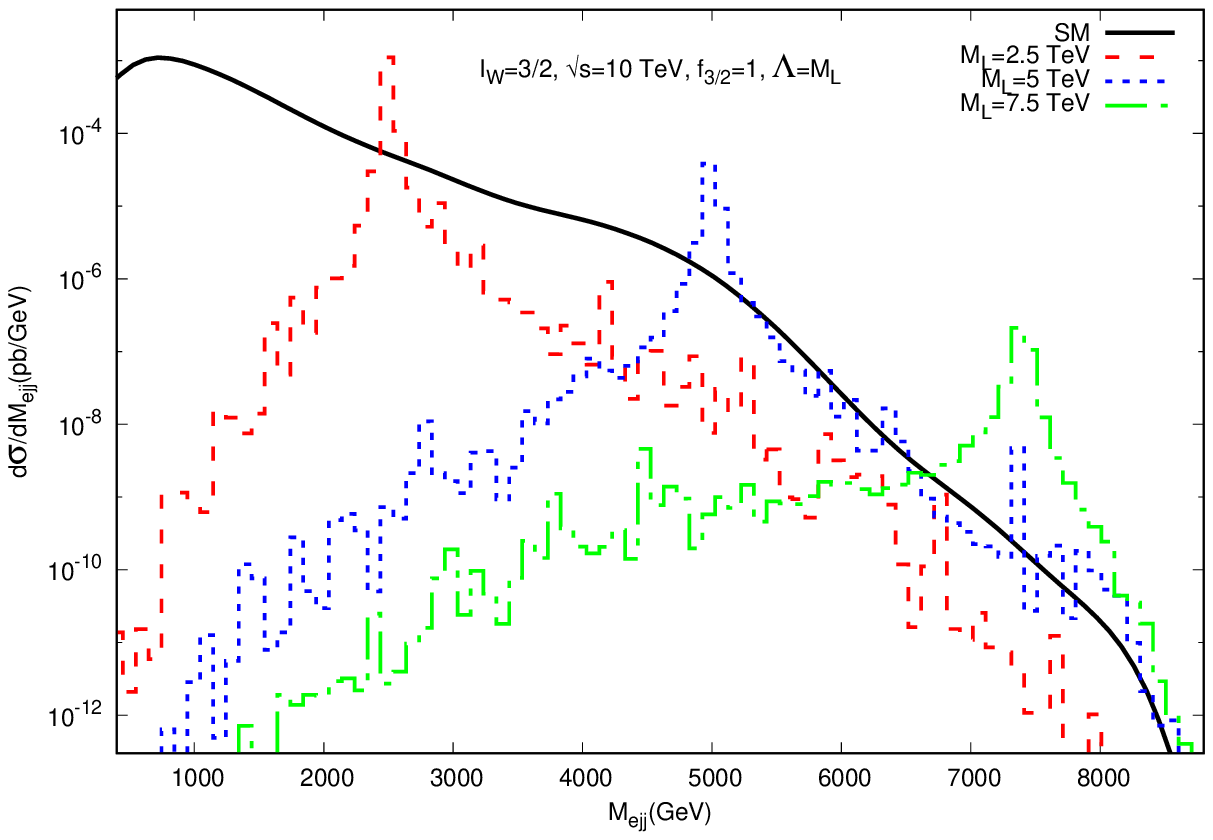}

\includegraphics[scale=0.7]{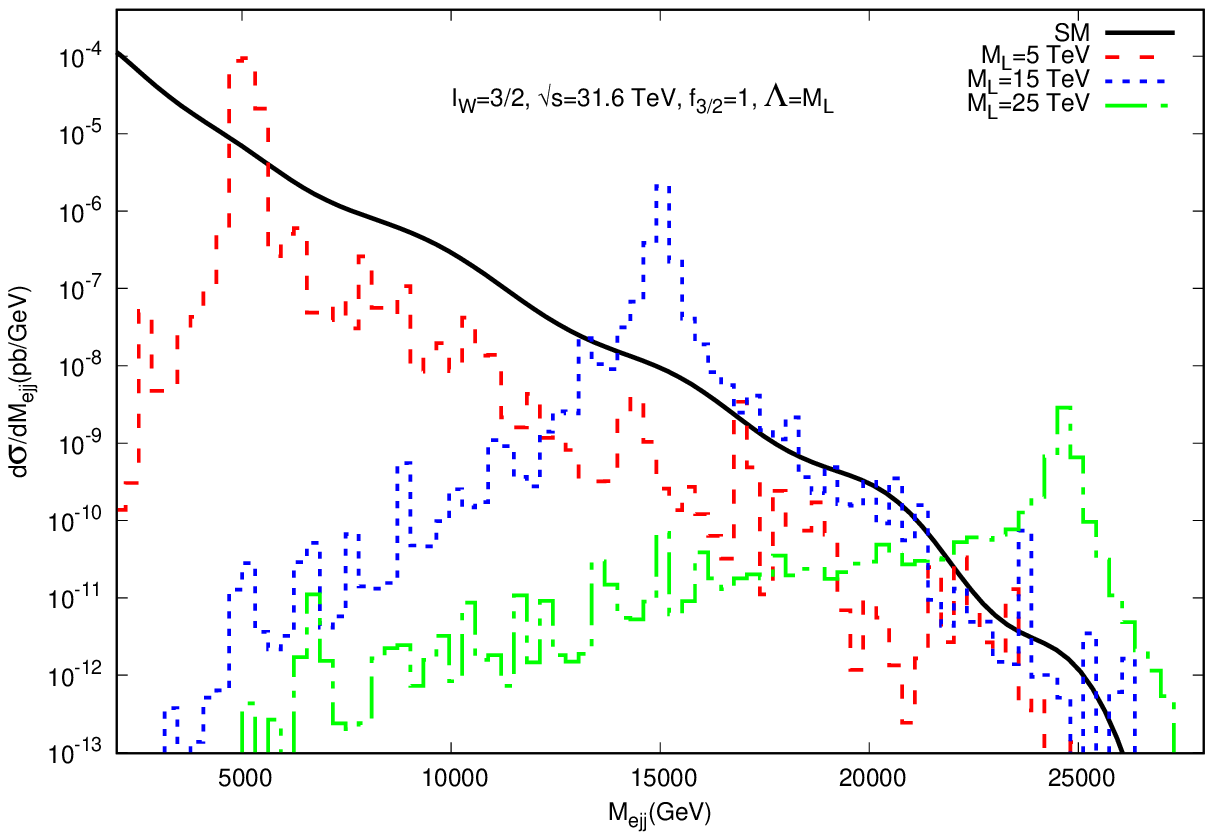}

\caption{Invariant mass distribution of $ejj$ system for $I_{W}=3/2$ after
the discovery cuts.}
\end{figure}

As expected, these distributions show a peak around
the chosen mass value of $L^{--}.$ Since we try to specify doubly
charged lepton signal from its decay products, we do not impose any
further cuts on jets. We define the discovery sensitivity
as

\[
SS=\frac{\left|\sigma{}_{S+B}-\sigma_{B}\right|}{\sqrt{\sigma_{B}}}\sqrt{L_{int}}
\].

Here, $\sigma_{S+B}$ is the cross section due to the presence of doubly
charged lepton, $\sigma_{B}$ is the SM background cross section,
and $L_{int}$ is the integrated luminosity of the collider. In Figures
10 and 11, we plot the $SS-M_{L}$ to determine the $2\sigma$ (exclusion),
$3\sigma$ (observation),and $5\sigma$ (discovery) limits.  

\begin{figure}[!htb]
\includegraphics[scale=0.7]{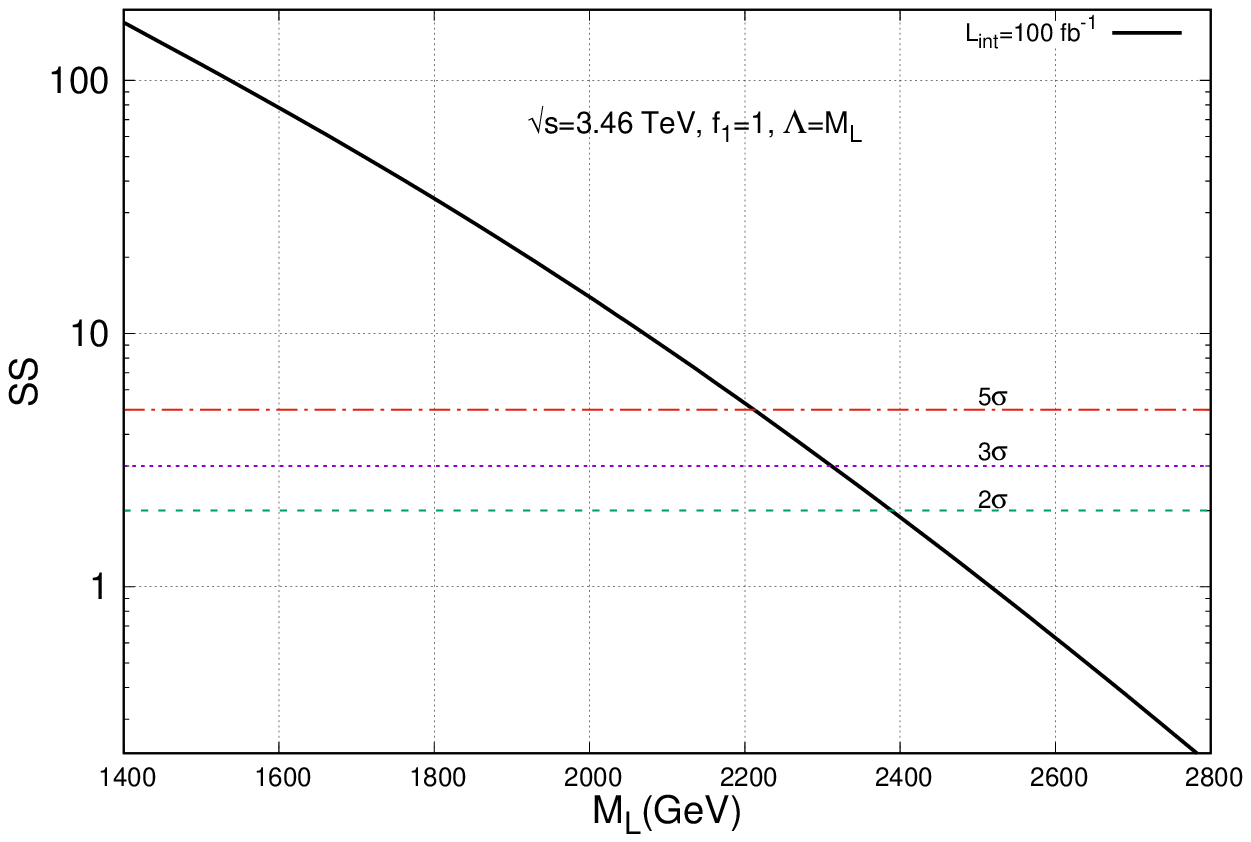}\includegraphics[scale=0.7]{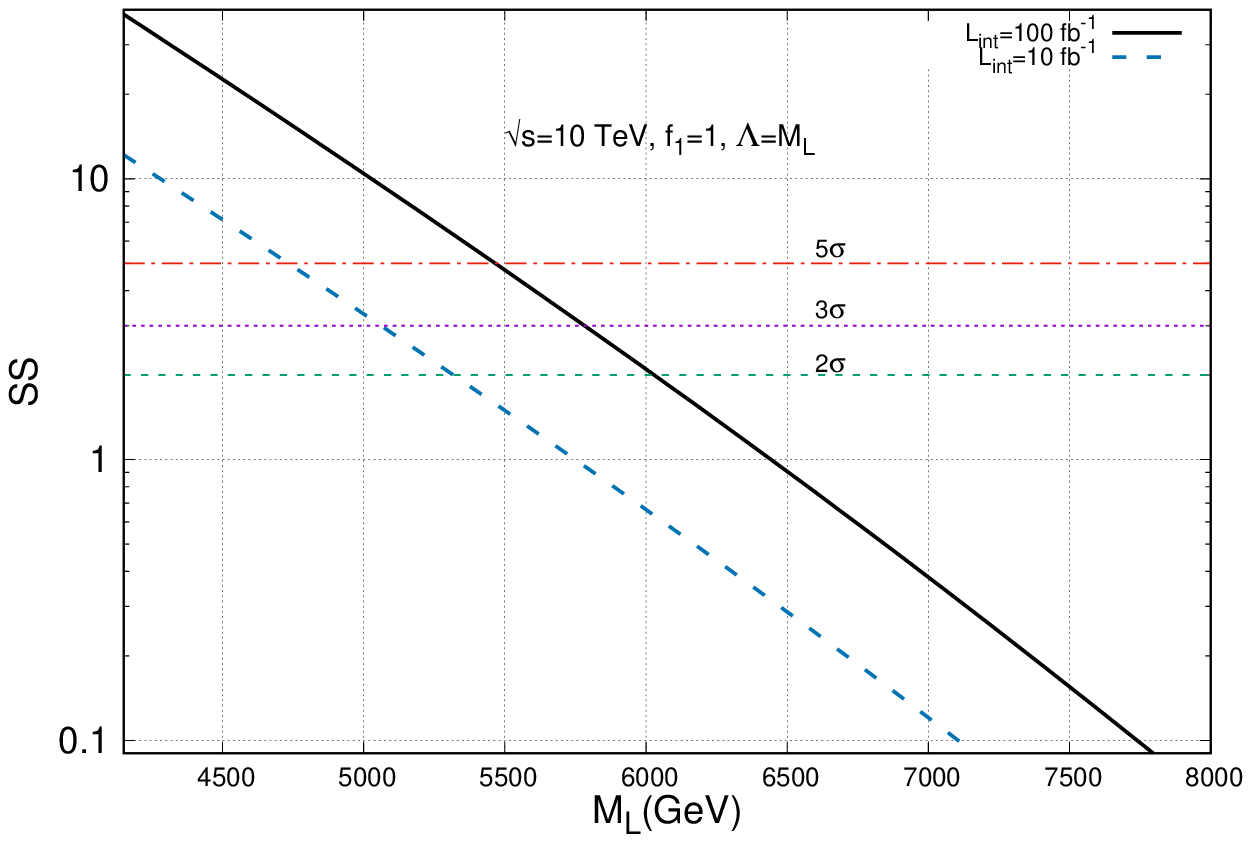} 

\includegraphics[scale=0.7]{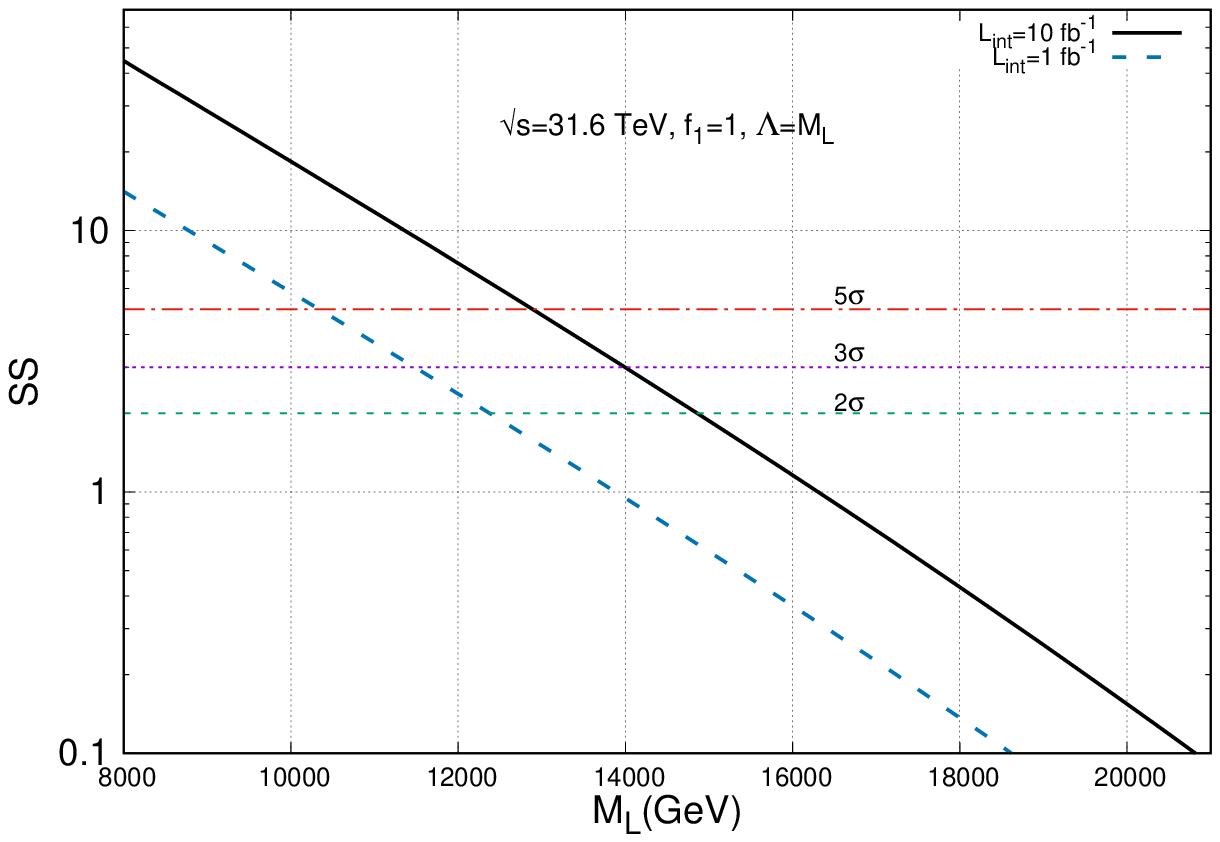}

\caption{$SS-M_{L}$ graphics for $I_{W}=1$ multiplet.}
\end{figure}

\begin{figure}[!htb]
\includegraphics[scale=0.7]{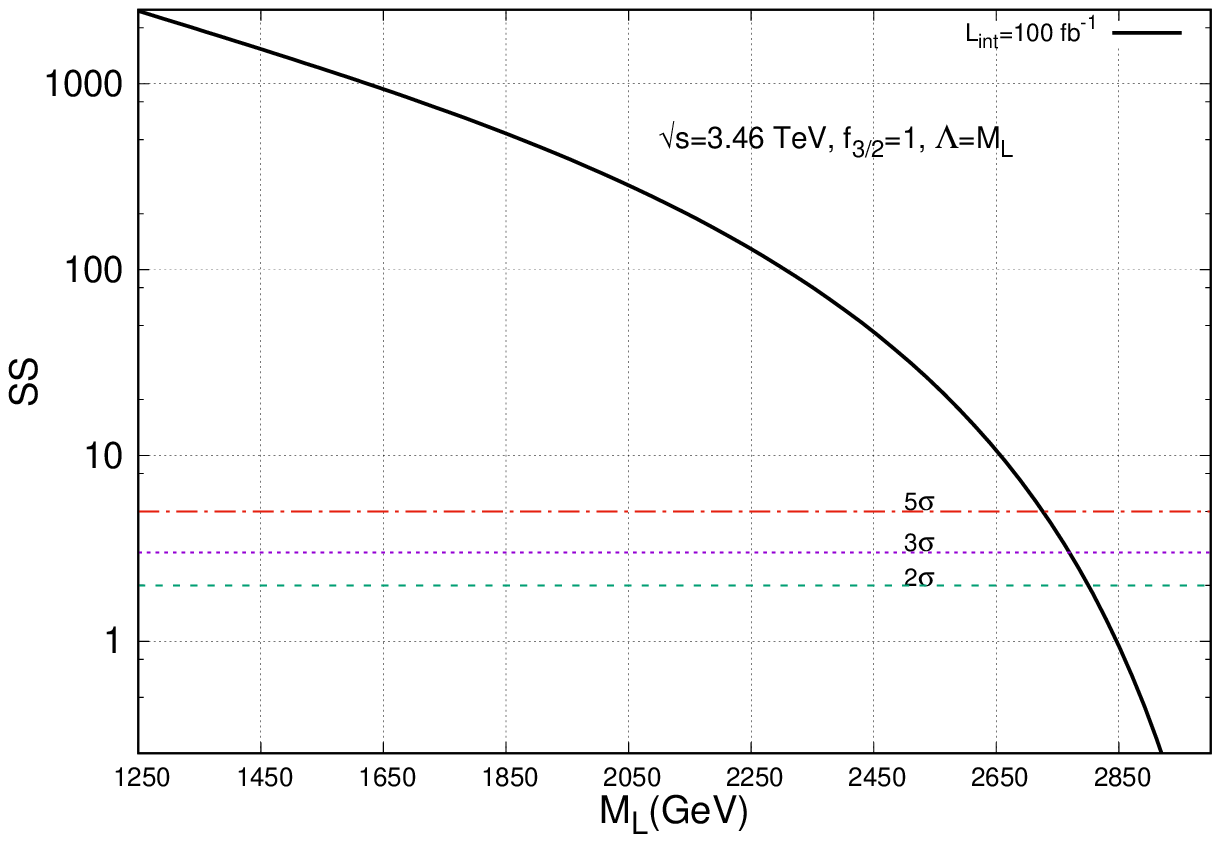}\includegraphics[scale=0.7]{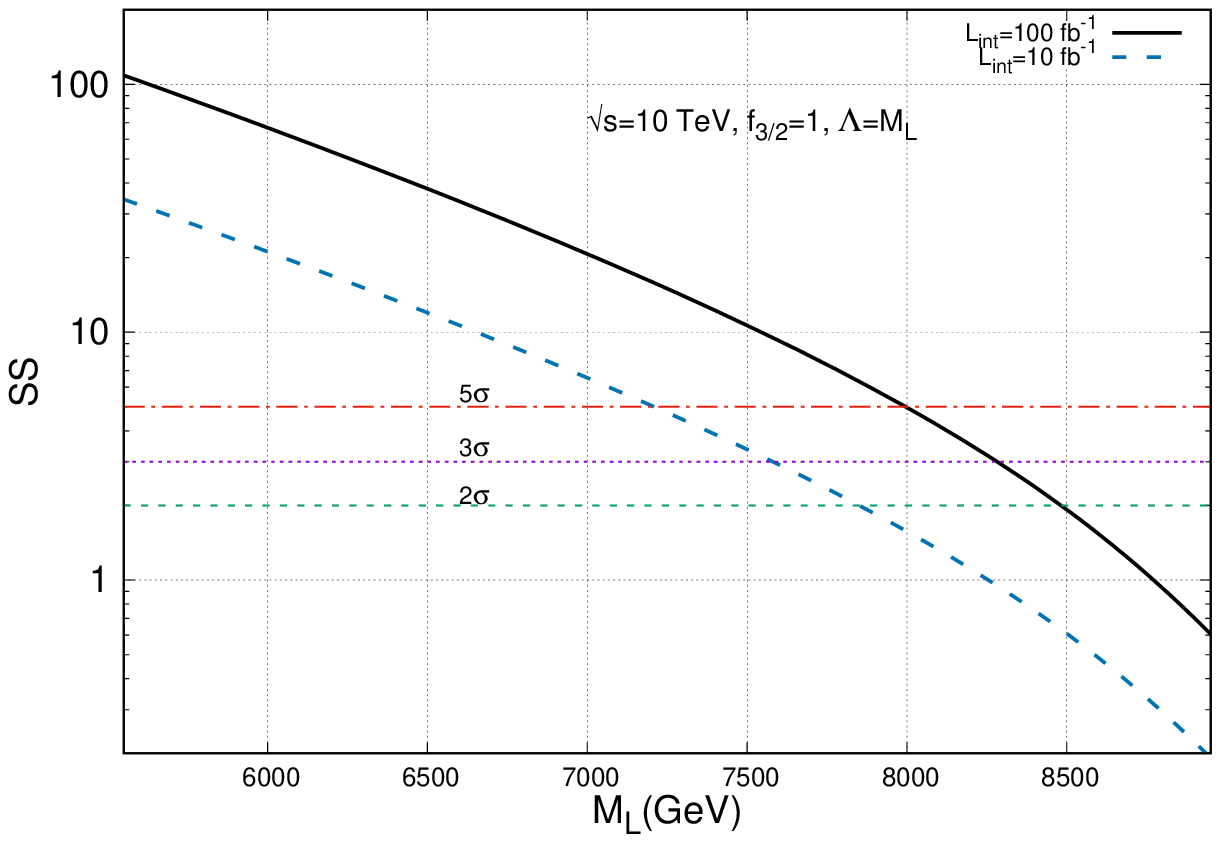}

\includegraphics[scale=0.7]{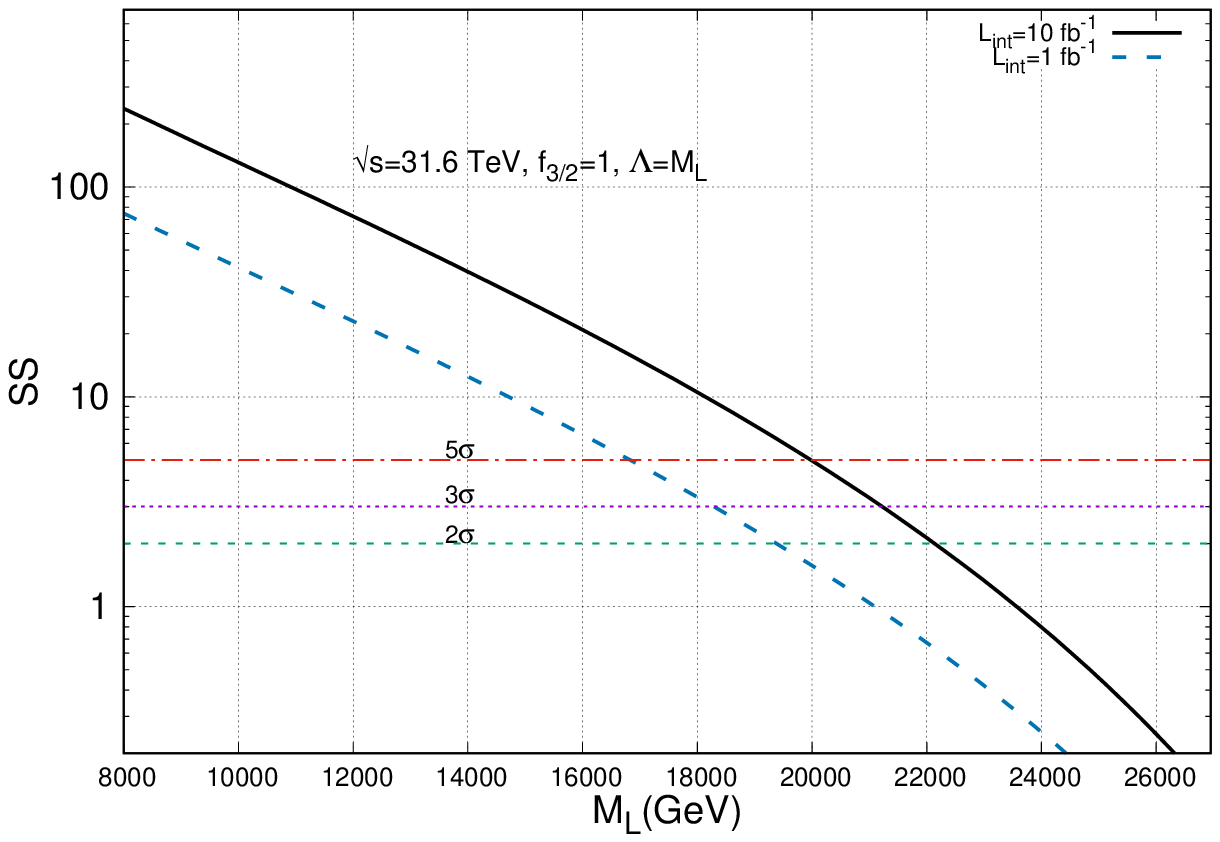}

\caption{$SS-M_{L}$ graphics for $I_{W}=3/2$ multiplet.}
\end{figure}

In Table IV, we give
the doubly charged lepton mass limits at different FCC-based $ep$
colliders for taking into account $f_{1}=f_{3/2}=1$ and $\Lambda=M_{L}$
concerning the criterias $SS>2$, $SS>3$, and $SS>5$ which denote
the exclusion, observation and discovery mass limits, respectively.

\begin{table}[!htb]

\caption{Mass limits for the doubly charged leptons for the FCC-based $ep$
colliders taking into account $I_{W}=1\,(I_{W}=3/2)$. }

\begin{tabular}{|c|c|c|c|c|c|}
\hline 
Collider & $\sqrt{s}$ (TeV)  & $L_{int}$($fb^{-1}$per year) & $2\sigma$ (TeV) & $3\sigma$ (TeV) & $5\sigma$ (TeV) \tabularnewline
\hline 
\hline 
ERL60$\otimes$FCC  & $3.46$ & $100$ & $2.38 (2.80) $ & $ 2.30 (2.77) $ & $ 2.21 (2.73)$\tabularnewline
\hline 
\multirow{2}{*}{ILC$\otimes$FCC } & \multirow{2}{*}{$10$} & $10$ & $5.33 (7.20)$ & $5.08 (7.56) $ & $4.74 (7.85)$\tabularnewline
\cline{3-6} 
 &  & $100$ & $6.02 (7.99)$ & $5.77 (8.28) $ & $5.46 (8.47)$\tabularnewline
\hline 
\multirow{2}{*}{PWFA-LC$\otimes$FCC } & \multirow{2}{*}{$31.6$} & $1$ & $12.4 (19.4) $ & $11.5 (18.3) $ & $10.3 (16.8)$\tabularnewline
\cline{3-6} 
 &  & $10$ & $14.9 (22.1)$ & $13.9 (21.2) $ & $12.9 (20.0)$\tabularnewline
\hline 
\end{tabular}
\end{table}

\section{conclus\i on}

A distinct and exclusive point of view of the compositeness is weak
isospin invariance. It enables us to extend the weak isospin values
to $I_{W}=1$ (triplet) and $I_{W}=3/2$ (quadruplet) multiplets.
Doubly charged leptons that have electrical charge of $Q=-2e$ appear
in these exotic multiplets. To find a clue about such new particles
at future high-energy and high-luminosity colliders that would indicate
the internal structure of the known fermions, we have presented a phenomenological
cut-based study for probing the doubly charged leptons coming from
extended weak isopin multiplets at various FCC-based $ep$ colliders.
Taking into consideration the lepton flavor conservation, we have dealt
with the decay of $L^{--}$ as $L^{--}\rightarrow e^{-}W^{-}$ and
$W$ boson as $W\rightarrow jj$ after the single production of doubly
charged lepton at $ep$ colliders. We have provided the $2\sigma$
, $3\sigma$ ,and $5\sigma$ statistical significance ($SS$) exclusion
curves in the $SS-M_{L}$ parameter space. Taking into criteria $SS > 5$ that corresponds to discovery, we have obtained the mass limits for doubly charged lepton for the exotic multiplet
$I_{W}=1$ ($I_{W}=3/2$), $2.21\,(2.73)$ TeV, $5.46\,(8.47)$ TeV, and $12.9\,(20.0)$TeV
at $\sqrt{s}=3.46$ TeV, $\sqrt{s}=10$ TeV, and $\sqrt{s}=31.6$
TeV, respectively. Our study has showed that FCC-based $ep$ colliders have quite well potential to attain the signals of doubly charged leptons considered in extended weak isospin models.

\subsection{Data Availability}
No data were used to support this study.

\end{document}